\documentclass[aps,showpacs,prr,twocolumn,superscriptaddress,floatfix]{revtex4-2}
 \usepackage{graphicx,bm,amssymb,amsmath,dcolumn,hyperref}
 \usepackage{multirow}
 \usepackage{color}
 \usepackage{subfigure}  
 \usepackage{epstopdf}
 \usepackage{bbm}
 \usepackage{graphicx}
 \usepackage{hyperref}
 \usepackage{threeparttable}
 \usepackage{amsthm,enumitem,mathrsfs}
 \usepackage[marginal]{footmisc}

\begin{document}

\title{On Sak's criterion for statistical models with long-range interaction}

\author{Tianning Xiao}
\thanks{These two authors contributed equally to this work}
\affiliation{Hefei National Research Center for Physical Sciences at the Microscale and School of Physical Sciences, University of Science and Technology of China, Hefei 230026, China}

\author{Ziyu Liu}
\thanks{These two authors contributed equally to this work}
\affiliation{Hefei National Research Center for Physical Sciences at the Microscale and School of Physical Sciences, University of Science and Technology of China, Hefei 230026, China}

\author{Zhijie Fan}
\email{zfanac@ustc.edu.cn}
\affiliation{Hefei National Research Center for Physical Sciences at the Microscale and School of Physical Sciences, University of Science and Technology of China, Hefei 230026, China}
\affiliation{Hefei National Laboratory, University of Science and Technology of China, Hefei 230088, China}
\affiliation{Shanghai Research Center for Quantum Science and CAS Center for Excellence in Quantum Information and Quantum Physics, University of Science and Technology of China, Shanghai 201315, China}

\author{Youjin Deng}
\email{yjdeng@ustc.edu.cn}
\affiliation{Hefei National Research Center for Physical Sciences at the Microscale and School of Physical Sciences, University of Science and Technology of China, Hefei 230026, China}
\affiliation{Hefei National Laboratory, University of Science and Technology of China, Hefei 230088, China}
\affiliation{Shanghai Research Center for Quantum Science and CAS Center for Excellence in Quantum Information and Quantum Physics, University of Science and Technology of China, Shanghai 201315, China}

\begin{abstract}
Determining the threshold value $\sigma_*$ that separates the short-range (SR) and long-range (LR) universality classes in phase transitions remains a controversial issue. While Sak's criterion, $\sigma_* = 2 - \eta_{\mathrm{SR}}$, has been widely accepted, recent studies of two-dimensional (2D) models with long-range interactions have challenged it.  
In this work, we focus on the crossover between LR and SR criticality in several classical 2D statistical models, including the XY, Heisenberg, percolation, and Ising models, whose interactions decay as $1/r^{2+\sigma}$. Our previous simulations for the XY, Heisenberg, and percolation models consistently indicate a universal boundary at $\sigma_* = 2$. Here, we complete the picture by performing large-scale Monte Carlo simulations of the 2D LR-Ising model, reaching lattice sizes up to $L = 8192$.  
By analyzing the Fortuin-Kasteleyn critical polynomial $R_p$, the Binder ratio $Q_m$, and the anomalous dimension $\eta$, we obtain convergent and self-consistent evidence that the universality class already changes sharply at $\sigma = 2$. Taken together, these results establish a unified scenario for LR interacting systems: across all studied models, the crossover from LR to SR universality occurs at $\sigma_* = 2$.

\end{abstract}
 

\maketitle


\section{Introduction}
\label{sec:intro}

Long-range (LR) interactions are ubiquitous in physical systems, ranging from dipolar forces in magnetic materials to gravitational interactions in astrophysical systems~\cite{campa2014}.
These interactions, typically decaying as $1/r^{d+\sigma}$ where $d$ is the spatial dimension and $\sigma$ controls the interaction range, fundamentally alter critical behaviors compared to their short-range (SR) counterparts~\cite{spivak2004, lahaye2009, peter2012}.
Understanding how LR interactions modify critical phenomena, therefore, remains a central question in statistical and condensed-matter physics.
Moreover, recent advances in both theoretical approaches~\cite{cpl_42_7_070002, yao2024nonclassical,2dlrheisen, 2dlrperco} and experimental platforms such as trapped ions and Rydberg atom arrays~\cite{lu2012, schauss2012, firstenberg2013, yan2013, aikawa2012, lewis2023} have further stimulated intensive research on this topic.

Despite much progress, several fundamental questions remain unresolved~\cite{defenu2023}. Among these, the most debated issue concerns the threshold value of $\sigma$ that separates the LR and SR universality classes. Specifically, for interactions decaying as $1/r^{d+\sigma}$, there exists a critical value $\sigma_*$ above which the system's critical behavior shares the same set of critical exponents as its SR counterpart. In contrast, below $\sigma_*$ the system belongs to distinct LR universality classes. The precise value of this threshold has been debated for decades~\cite{fisher1972, sak1973, Luijten2002, picco2012, angelini2014, horita2017, cpl_42_7_070002, yao2024nonclassical, 2dlrheisen, PhysRevB.107.224204}.

The first systematic investigation of this problem was carried out by Fisher \textit{et al.}~\cite{fisher1972}, who employed a second-order $\epsilon$-expansion technique to analyze a field-theoretical description of the LR-O($n$) spin model. This model extends the Ising spin ($s=\pm 1$) to an $n$-component unit vector, with $n = 1, 2, 3$ corresponding to the Ising, XY, and Heisenberg models, respectively. In their work, they proposed three distinct regimes based on the decay exponent $\sigma$:
(1) Classical regime  ($0 < \sigma < d/2$): The critical behavior in this regime is governed by mean-field theory, characterized by Gaussian fixed points;
(2) Nonclassical regime  ($d/2 < \sigma < 2$): Gaussian fixed points become unstable, and the critical exponents acquire $\sigma$-dependence. Nonetheless, the anomalous magnetic dimension $\eta$ was conjectured to retain its mean-field value: $\eta = 2 - \sigma$;
(3) SR regime ($\sigma > 2$): The LR interactions become irrelevant, and the system exhibits SR critical behavior.
According to their analysis, the boundary separating the nonclassical and SR regimes is located at $\sigma_* = 2$ for all LR-O($n$) models.

\begin{figure}
    \centering
    \includegraphics[width=\linewidth]{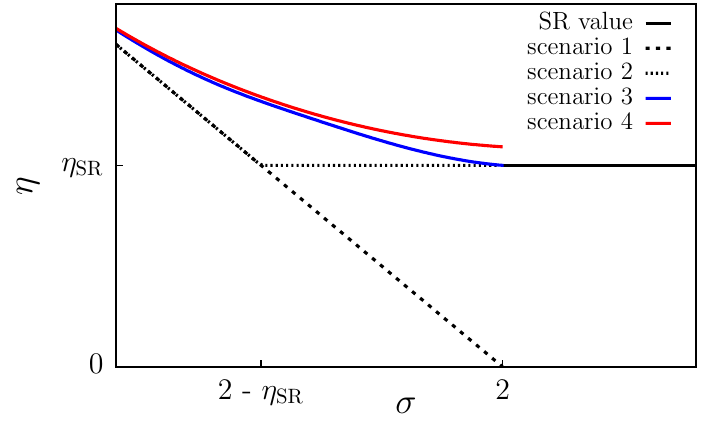}
    \caption{Sketch of the anomalous dimension $\eta$ as a function of $\sigma$ for four scenarios. 
    For $\sigma > 2$, it is uncontroversial that the value belongs to the SR case, as indicated by the black solid line.  
    As for $\sigma \leq 2$, (i) the prediction by Fisher, $\eta = 2 - \sigma$~\cite{fisher1972}, is shown with a sparse dashed line. A jump in $\eta$ occurs at $\sigma = 2$, marking the transition point of the universality class at $\sigma_* = 2$.  
    (ii) Sak's criterion, $\eta = \max(2 - \sigma, \eta_{\rm SR})$~\cite{sak1973}, is shown with a dense dashed line. Here the transition point of the universality class is at $\sigma_* = 2 - \eta_{\rm SR}$ (for the Ising model, $\eta_{\rm SR} = 1/4$).  
    (iii) Picco proposed a third scenario based on his numerical results~\cite{picco2012}, where the transition point is at $\sigma_* = 2$, smoothly connecting to the short-range region, shown by the blue solid line.  
    (iv) We propose a fourth scenario, where although the curve smoothly connects to $\sigma = 2$, there is still a jump at $\sigma = 2$, indicating that the nature of the phase transition at $\sigma = 2$ is already different from the SR case, shown by the red solid line.}
    \label{fig:our_scenario}
\end{figure}

However, this scenario implied a discontinuity in the anomalous magnetic dimension $\eta$ at $\sigma = 2$, where $\eta$ would abruptly jump from $0$ to the SR model's anomalous magnetic dimension $\eta_{\mathrm{SR}}$, as shown by the sparse dashed line in Fig.~\ref{fig:our_scenario}. While such a discontinuity is not necessarily pathological from the perspective of universality and renormalization group (RG) theory~\cite{blumeTheoryFirstOrderMagnetic1966,capelPossibilityFirstorderPhase1966,blumeIsingModelTransition1971}, Sak proposed a revised scenario: $\sigma_* = 2 - \eta_{\mathrm{SR}}$~\cite{sak1973}, which leads to $\eta = \max(2 - \sigma, \eta_{\mathrm{SR}})$. Sak's prediction thus removes the discontinuity of $\eta$ during the crossover between the LR and SR regimes, as shown by the dense dashed line in Fig.~\ref{fig:our_scenario}. 
Sak further suggested that the continuous connection between the two regimes, observed at leading order in the $\varepsilon$-expansion, might persist to all perturbative orders.
This refinement, now known as Sak's criterion, has been widely adopted in theoretical studies of LR systems~\cite{defenu2023}. Nonetheless, the issue remains debated, with differing viewpoints and no clear consensus in the field-theoretical treatment~\cite{YAMAZAKI1977207, YAMAZAKI1978446, vanenter1982,blanchard2013}.

From a numerical perspective, the situation is equally challenging and marked by longstanding controversies. In a pioneering work, Luijten and Blöte developed a specialized cluster Monte Carlo (MC) algorithm~\cite{luijten1995} for the two-dimensional (2D) LR-Ising model ($\eta_{\rm SR} = 1/4$) to investigate the model. By analyzing critical exponents and Binder cumulants~\cite{Luijten2002}, they found the threshold at $\sigma_* = 7/4$ supporting Sak's criterion. However, their data exhibited significant statistical uncertainties near $\sigma = 2$ in the 2D case. This issue was later highlighted by Picco~\cite{picco2012}, who questioned the reliability of Luijten's conclusions based on the results of the improved numerical techniques and instead proposed a third alternative scenario: As $\sigma \approx 2$, the anomalous dimension $\eta$ smoothly varies from the LR values to the SR value $\eta_{\mathrm{SR}}$, as shown in the blue line of Fig.~\ref{fig:our_scenario}. Blanchard \textit{et al.} further supported this interpretation with complementary field-theoretical arguments~\cite{blanchard2013}.
However, the debate persisted when Angelini \textit{et al.} \cite{angelini2014} challenged Picco's results from a finite-size scaling (FSS) perspective, employing a double-power scaling approach that led them to question the smooth crossover scenario. Subsequently, Horita \textit{et al.}\cite{horita2017} introduced the method of self-combined ratios, designed to suppress finite-size corrections more effectively. 
Their results appeared to support Sak's criterion once again; however, this technique has a flaw that can potentially mask genuine universal features, as it could artificially compress the difference in the universal values between the nonclassical and SR regimes (see Section~\ref{subsec:Rp} for a detailed discussion).

Very recently, progress has been made in the study of continuous spin models (O($n$) spin models with $n \geq 2$).  
The 2D LR dilute XY model~\cite{PhysRevB.100.054203} and the standard 2D LR-XY model~\cite{cpl_42_7_070002,yao2024nonclassical} consistently indicate that $\sigma_* = 2$, thus contradicting Sak's prediction.  
Meanwhile, these studies did not observe the two phase transitions predicted by Giachetti \textit{et al.}~\cite{giachetti2021,giachetti2022}.  
Moreover, the 2D LR-Heisenberg model, which exhibits long-range order (LRO) for $\sigma \leq 2$~\cite{2dlrheisen}, also supports $\sigma_* = 2$.  
Notably, the universal dimensionless ratios and several critical exponents of the 2D LR-percolation model~\cite{2dlrperco, grassberger2013}, including the anomalous dimension $\eta$, the correlation-length exponent $\nu$ and the shortest-path exponent $d_{\rm min}$, exhibit a clear distinction between $\sigma > 2$ and $\sigma \leq 2$, and a discontinuity at $\sigma = 2$ has also been observed.  
These consistent findings across different models strongly suggest $\sigma_* = 2$, motivating a thorough re-examination of the 2D long-range Ising model, which has historically been the central case of this controversy.

\begin{figure*}[!ht]
    \centering
    \includegraphics[width=\linewidth]{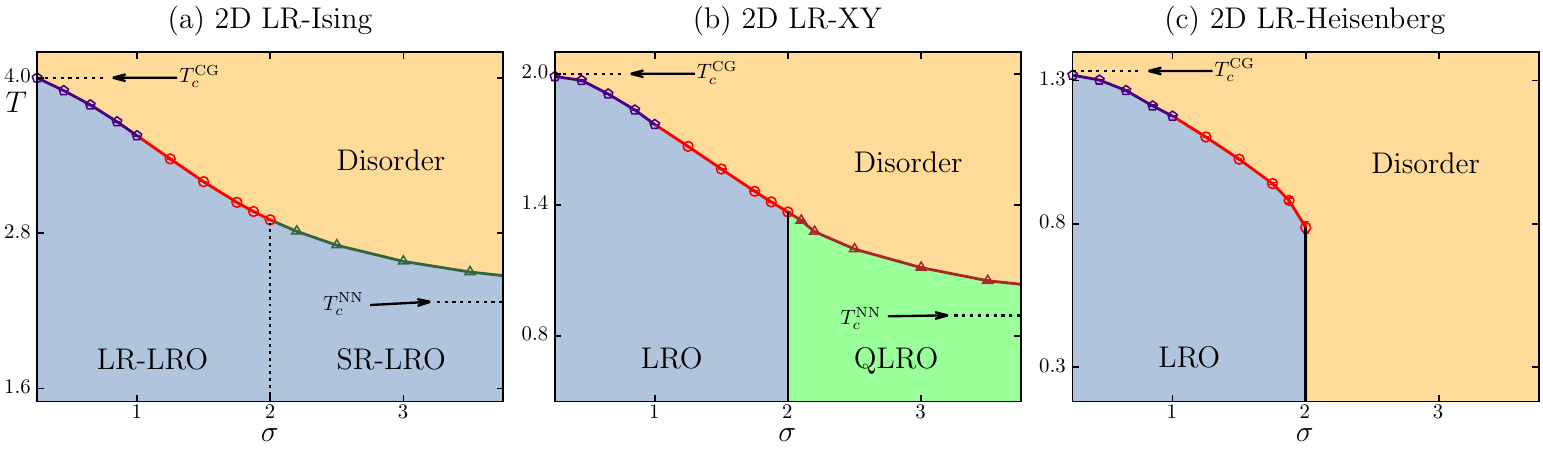}
    \caption{Phase diagrams of the 2D O($n$) spin models: Ising ($n=1$), XY ($n=2$)~\cite{cpl_42_7_070002,yao2024nonclassical}, and Heisenberg ($n=3$)~\cite{2dlrheisen}, with long-range (LR) interactions decaying as $1/r^{2+\sigma}$. In all panels, $T_c^{\text{CG}} = 4/n$ denotes the complete-graph critical temperature~\cite{kirkpatrick2017} under the normalization in Eq.~\eqref{eq:norm}, and $T_c^{\text{NN}}$ indicates the NN limit~\cite{PhysRev.65.117, komura2012a}. The classical regime ($\sigma \leq 1$, purple line) follows mean-field behavior.  
    (a) For the Ising case, the system exhibits a second-order transition throughout. The universality class changes at $1 < \sigma \leq 2$ (red line), and for $\sigma > 2$ (green line), it recovers the SR Ising class. The dashed line indicates that the geometric properties of the low-temperature phase differ between $\sigma < 2$ and $\sigma > 2$, in the FK representation of the Ising model. 
    (b) For the XY model, the SR regime ($\sigma > 2$, brown line) undergoes a Bereinzinskii-Koterlitz-Thouless (BKT) topological phase transition into a quasi-long-range-order (QLRO) phase, while for $1 < \sigma \leq 2$, a direct second-order transition into an LRO phase appears.  
    (c) For the Heisenberg model, finite-temperature transitions exist only for $\sigma \leq 2$.  
    All three models consistently exhibit a universality boundary at $\sigma_* = 2$.
    }
    \label{fig:PD}
\end{figure*}

In this work, we provide a unified investigation of the crossover between LR and SR criticality in various 2D statistical models, including the XY, Heisenberg, percolation, and Ising models, with interactions decaying as $1/r^{2+\sigma}$.
Building upon our studies on the XY, Heisenberg, and percolation models -- each of which consistently suggested a universality boundary at $\sigma_* = 2$, we now complete the picture by performing large-scale MC simulations of the 2D LR-Ising model.
We employ an enhanced variant of the Luijten–Blöte cluster algorithm~\cite{luijten1995, luijten1997, michel2019}, enabling simulations up to $L = 8192$.
In particular, we analyze the Ising model from a geometric perspective by measuring the Fortuin–Kasteleyn (FK) critical polynomial $R_p$, whose universal value is exactly known for the SR Ising universality class~\cite{Jacobsen_2014}.
These geometric observables, together with the Binder ratio $Q_m$ and the anomalous dimension $\eta$, allow us to determine the universality boundary with high precision and demonstrate that the crossover from LR to SR criticality occurs sharply at $\sigma = 2$.

The remainder of this paper is organized as follows. 
Firstly, in Sec.~\ref{sec:review}, we review our previous studies on the 2D XY~\cite{cpl_42_7_070002,yao2024nonclassical}, Heisenberg~\cite{2dlrheisen}, and percolation~\cite{2dlrperco} models, which together provide an overview of 2D statistical systems with LR interactions.
After that, in Sec.~\ref{sec:results}, we focus on the 2D LR-Ising model and determine its critical points for various $\sigma$ by measuring both the FK critical polynomial $R_p$ and the Binder ratio $Q_m$.
These two observables yield consistent estimates within error bars, as summarized in the phase diagram shown in Fig.~\ref{fig:PD}(a).
By performing crossing-point extrapolations, supplemented with least-squares fitting, we obtain the universal values of these observables at criticality, which reveal whether the universality class changes across $\sigma$.
Furthermore, the critical exponents are extracted from the FSS of the magnetic susceptibility $\chi$ and the scaled covariance $g^{(x)}_{ER}$, respectively.
The resulting quantities, listed in Table~\ref{tab:summary}, consistently demonstrate that the crossover between the LR and SR regimes occurs precisely at $\sigma = 2$.
In particular, both $R_p$ and the anomalous dimension $\eta$ exhibit clear discontinuities at $\sigma = 2$, in agreement with the behavior of the LR-percolation model~\cite{2dlrperco}. These findings, shown in Figs.~\ref{fig:PD_Rp} and~\ref{fig:eta_yt}, also stand in contrast to earlier numerical studies supporting Sak's criterion~\cite{Luijten2002, horita2017}.
Therefore, in Sec.~\ref{sec:scenario}, we propose a fourth scenario: although the critical exponents vary smoothly within the LR regime, the crossover at $\sigma_* = 2$ remains sharp and discontinuous, as illustrated by the red line in Fig.~\ref{fig:our_scenario}.
Finally, Sec.~\ref{sec:discussion} concludes the paper and discusses the broader implications for 2D LR statistical models.

\section{Overview of the 2D statistical models with long-range interaction}
\label{sec:review}

To provide an overview of our findings, Fig.~\ref{fig:PD} displays the phase diagrams of 2D LR-O($n$) spin models: Ising ($n=1$), XY ($n=2$)~\cite{cpl_42_7_070002,yao2024nonclassical}, and Heisenberg ($n=3$)~\cite{2dlrheisen} models. Furthermore, the phase diagram of the LR-percolation model~\cite{2dlrperco} shares the same topology as the LR-Ising case.
Also, the phase diagram of the 2D LR uniform forest (UF) model~\cite{2dlrheisen}, which is a graphical model but can be mapped to a non-Gaussian fermionic theory with non-abelian continuous OSP($1|2$) supersymmetry by generalizing Kirchhoff's matrix-tree theorem~\cite{PhysRevLett.93.080601}, has the same topology as that of the Heisenberg model.
Remarkably, all these models consistently exhibit a universality boundary at $\sigma_* = 2$, signaling the crossover between the SR and LR universality classes.

Here, the Hamiltonian of the 2D LR-O($n$) model on a square lattice with periodic boundary conditions (PBCs), i.e., on a torus $\mathbb{T}^2$, is defined as
\begin{align}
    \mathcal{H} = - \sum_{i<j}^{N} \frac{c(\sigma, L)}{r_{ij}^{2+\sigma}}\, \mathbf{S}_i \cdot \mathbf{S}_j,
    \label{eq:LRON}
\end{align}
where $\mathbf{S}_i$ denotes an $n$-component unit vector at site $i$, and $N = L \times L$ is the total number of sites. 
The interactions follow the shortest path on its surface, corresponding to the minimum-image convention~\cite{frenkel_understanding_2002,christiansen_phase_2019, agrawal_kinetics_2021}.
Hence, the summation in Eq.~\eqref{eq:LRON} runs over all spin pairs, resulting in $N(N-1)/2$ interaction terms. The normalization constant $c(\sigma, L)$ is chosen to satisfy
\begin{align}
    \sum_{j=2}^{N} \frac{c(\sigma, L)}{r_{1j}^{2+\sigma}} = 4,
    \label{eq:norm}
\end{align}
where the value $4$ corresponds to the number of nearest-neighbor (NN) sites.
This choice ensures that the model continuously interpolates between the NN limit ($\sigma \rightarrow \infty$) and the complete-graph (CG) limit ($\sigma \rightarrow -2$) with critical temperature $T_c^{\text{CG}} = 4/n$~\cite{kirkpatrick2017}. This normalization effectively mitigates finite-size corrections and is asymptotically equivalent to the Ewald summation treatment~\cite{frenkel2002,janke2019,agrawal2021}. Importantly, it only rescales the temperature scale and does not alter the physical properties of the model.

Similarly, the bond probability of the 2D LR bond percolation model between site $i$ and site $j$ is defined as
\begin{align}
p(r_{ij}) = \frac{K}{r_{ij}^{2+\sigma}},
\label{eq:LRP}
\end{align}
where the coupling strength $K \in (0,1]$. In the limit $\sigma \rightarrow \infty$, the bond probability reduces to a constant $K$ for NN bonds and vanishes for all other types of bonds, reducing the model to the NN case. As $\sigma$ decreases, the probability of forming LR bonds gradually increases. In the limit $\sigma \rightarrow -2$, the bond probability approaches a constant $K$ for all types of bonds, and the percolation model reduces to the CG case.

With the Hamiltonian defined, we now briefly discuss the key features of these 2D LR models.


For the continuous spin models ($n \ge 2$), the universality can be diagnosed from the intrinsic nature of their phases. As shown in Fig.~\ref{fig:PD}(b) and (c), due to the Mermin–Wagner theorem~\cite{mermin_absence_1967}, the LRO phase does not exist in the SR case, but appears in the presence of sufficient LR interaction. The rationale is straightforward: the observation of LRO implies that the universality cannot belong to the SR class, unless an extremely nontrivial mechanism intervenes.  
In the 2D LR-XY~\cite{yao2024nonclassical} and Heisenberg~\cite{2dlrheisen} models, the systems develop clear LRO through a single phase transition for $\sigma \leq 2$, as evidenced by a finite square magnetization density $\langle M^2 \rangle > 0$ in the limit $L \to \infty$, where $M = L^{-2} \left| \sum_i \mathbf{S}_i \right|$. To mitigate finite-size corrections and make the existence of LRO more visible, we define the residual squared magnetization as
\begin{align}
    M_r^2 = \langle M^2 \rangle - b \langle M_k^2 \rangle ,
\end{align}
where $M_k = L^{-2} |\sum_i \mathbf{S}_i e^{i\mathbf{k}\cdot \mathbf{r}_i}|$ is the Fourier mode of magnetization, and $b > 0$ is a tuning parameter obtained from fitting. Here, $\langle \cdot \rangle$ denotes the statistical average. Since $M_k^2$ shares the same scaling as the leading term of $M^2$, but vanishes in the limit $L \to \infty$, $M_r^2$ therefore converges to the value of $M^2$ with significantly reduced finite-size effects. For $\sigma \leq 2$, we indeed observe that $M_r^2$ converges to a finite positive value as $L \to \infty$, with much smaller finite-size corrections. More importantly, since $b$ is positive, $M_r^2$ serves as a lower bound for $\langle M^2 \rangle$, and we already observe positive $M_r^2$ for large system sizes, which further confirms the presence of LRO for $\sigma \leq 2$.

Moreover, for the continuous spin models, the LRO phase breaks a continuous symmetry, giving rise to Goldstone modes. For $\sigma < 2$, the correlation function exhibits an algebraic tail, $g(x)\sim g_0 + x^{-\eta_{\ell}}$ with $\eta_{\ell} = 2 - \sigma$~\cite{yao2024nonclassical}, leading to the Fourier susceptibility $\chi_k = L^2 \langle M_k^2 \rangle  \sim L^{\sigma}$. This scaling of the Fourier susceptibility is the key signature of the LR Goldstone phase, which is clearly observed in~\cite{yao2024nonclassical}.  
For $\sigma = 2$, the logarithmic scaling $\chi_k \sim L^2 / \ln (L/L_0)$ is observed; therefore, we conjecture that the correlation function follows $g(x) \sim g_0 + 1/\ln (x/x_0)$. Here, $g_0$ is the squared magnetization in the thermodynamic limit, $L_0$ and $x_0$ are non-universal parameters.

Notably, Ref.~\cite{PhysRevLett.87.137203} extends the Mermin–Wagner theorem to the LR case, predicting that LRO cannot exist at $\sigma = 2$ in 2D for the LR-XY and LR-Heisenberg models. However, our numerical results clearly reveal the presence of LRO and Goldstone modes at $\sigma = 2$. Similar behaviors are observed in our Lévy-flight simulations~\cite{2dlrheisen}, where the nonanalytic propagator $q^{\sigma}$ contributes to the Goldstone mode, and the logarithmic scaling at $\sigma = 2$ is also observed.  
It is also worth emphasizing that the original Mermin–Wagner theorem rules out the spontaneous breaking of continuous symmetry when the second moment of the interaction is finite, i.e., $\sum_{\bm r}{\bm{r}}^2 J(\bm{r}) < \infty$~\cite{mermin_absence_1967}. For the interaction $J(r) \propto 1/r^{2+\sigma}$ in 2D, this condition fails at $\sigma = 2$, where the summation diverges logarithmically.  
In any case, the discrepancy between our numerical results and existing theoretical arguments underscores that the nature of the marginal case at $\sigma=2$ deserves further theoretical investigation.


From another viewpoint, the universality can also be identified from the behavior of the correlation length $\xi$ as the temperature $T$ approaches the critical point $T_c$ from the disordered side. For convenience in MC measurements, the correlation length is defined through the second-moment estimator,
\begin{align}
\xi = \frac{1}{2 \sin(|\mathbf{k}|/2)} \sqrt{\frac{\langle M^2 \rangle}{\langle M^2_k \rangle} - 1}.
\end{align}  
For the 2D LR-XY model, it is crucial to distinguish between the Berezinskii–Kosterlitz–Thouless (BKT) transition, where $\xi$ diverges exponentially as $\xi \sim \exp(b/\sqrt{t})$, and the second-order transition, where $\xi \sim t^{-\nu}$, with $t = (T - T_c)/T_c$ and $\nu$ as the correlation length exponent. In our studies~\cite{cpl_42_7_070002,yao2024nonclassical}, we found that for $\sigma \le 2$, both direct observations and extrapolations of $\xi$ clearly follow a power-law divergence, confirming a genuine second-order transition, whereas for $\sigma > 2$, an exponential scaling of $\xi$ is demonstrated, illustrating a BKT transition.
A similar pattern emerges for the Heisenberg case~\cite{2dlrheisen}. For $\sigma \le 2$, $\xi$ exhibits a power-law divergence near some finite $T_c$ as, $\xi \sim t^{-\nu}$. For $\sigma > 2$, the scaling $\xi \sim \exp(b/T)$ is identified, where $b$ is a non-universal constant, which indicates the absence of finite-temperature transitions, consistent with asymptotic freedom in the 2D SR Heisenberg model.


Finally, from the viewpoint of critical exponents, our results are also consistent with the above observations. Both $\eta$ and $\nu$ exhibit a smooth evolution across the nonclassical regime. For small $\sigma$, $\eta$ follows the long-range prediction $\eta = 2 - \sigma$. However, as $\sigma$ approaches 2, $\eta$ decreases more slowly than $2-\sigma$ and shows a slight discontinuity at $\sigma = 2$. For $\sigma > 2$, $\eta$ approaches and saturates to the short-range value $\eta_{\mathrm{SR}}$. This deviation near $\sigma = 2$ provides another quantitative signature of the crossover in universality.  
Meanwhile, field-theoretical analyses~\cite{giachetti2021,giachetti2022} predict an additional QLRO phase in the XY case for $7/4 < \sigma \le 2$, but such an intermediate phase was not observed in our numerical studies~\cite{cpl_42_7_070002,yao2024nonclassical}.

In contrast, for models such as percolation and Ising ($n=1$), the situation is more subtle. Their phase transitions are second-order across all $\sigma$, and no qualitative distinction appears between the LR and SR universality classes in terms of their high- and low-temperature phases, as shown in Fig.~\ref{fig:PD}(a). Therefore, their universality must be inferred from critical properties rather than from the phase structure itself. This poses a significant numerical challenge, as strong finite-size corrections make it difficult to distinguish universal behaviors with precision.

To address this challenge, we consider the $q$-state Potts model~\cite{potts1952some} whose Hamiltonian is given by 
\begin{align}
    \mathcal{H} = - \sum_{i < j} K_{ij} \delta_{s_i, s_j},
    \label{eq:potts}
\end{align}
where $s_i = 1, 2, \dots, q$ is the Potts spin, and $\delta_{s_i, s_j}$ is the Kronecker delta, equal to $1$ if $s_i = s_j$ and $0$ otherwise.
For the 2D LR case, we set $K_{ij} = c(\sigma, L)/r_{ij}^{2+\sigma}$, similar to Eq.~\eqref{eq:LRON}.  
Notably, this model can be mapped to the FK representation~\cite{fortuin1972random,swendsen1987}, where the configuration consists of clusters, and its partition function can be written as  
\begin{align}
    \mathcal{Z}_{\rm FK} = \sum_{\mathcal{G}} \left( \prod_{(i,j)\in \mathcal{G}} u_{ij} \right) q^{N_c},
    \label{eq:FK}
\end{align}
where $u_{ij} = e^{K_{ij}} - 1$, and the graph $\mathcal{G}$ represents the FK bond variables. Each bond $(i,j)$ connects sites $i$ and $j$, and $N_c$ denotes the number of clusters.  
Note that percolation and the Ising model correspond to the $q \to 1$ and $q = 2$ Potts models, respectively. Therefore, the FK representation provides a geometric framework that allows both percolation and Ising models to be studied from a geometric perspective.

For the 2D $q$-state Potts model~\cite{Jacobsen_2014}, one can define a critical polynomial as 
\begin{align}
R_p = \langle \mathcal{R}_2 \rangle - q \langle \mathcal{R}_0 \rangle,
\end{align}
where $\langle \mathcal{R}_2 \rangle$ denotes the probability that a spanning cluster emerges to connect itself by wrapping around the periodic boundaries along both the $x$ and $y$ directions, and $\langle \mathcal{R}_0 \rangle$ represents the probability that no cluster wraps at all. 
At the critical point, both $\langle \mathcal{R}_2 \rangle$ and $\langle \mathcal{R}_0 \rangle$ take universal values, but $R_p=0$ provides a more convenient criterion. 
Owing to the duality of planar graphs~\cite{Jacobsen_2014}, this exact relation $R_p=0$ holds at the critical point of the NN model, without any finite-size corrections, for the $q$-state Potts model on the square or triangular lattices. 
Although an exact proof is absent for non-planar graphs, previous numerical studies~\cite{xu2021, PhysRevE.98.062101} have shown that $R_p=0$ remains valid for models belonging to the SR universality class.  
Therefore, deviations of $R_p$ from zero can serve as a sensitive probe for detecting the LR–SR crossover.

For the percolation case ($q=1$), in addition to the conventional method of generating percolation configurations for a given $K$, we also employ the event-based ensemble method~\cite{li_explosive_2023}. This method enables simulations directly at pseudo-critical points, thereby avoiding potential systematic errors arising from inaccurate critical-point determination. Utilizing both the conventional and event-based methods, we achieved system sizes up to $L = 16384$ and determined the universal values of various dimensionless quantities ($Q_m$ and $R_p$) and the critical exponents, including $\eta$, $y_t = 1/\nu$, and the shortest-path exponent $d_{\rm min}$~\cite{2dlrperco}, with high precision. The results reveal a pronounced jump in both the universal ratios and the critical exponents at $\sigma = 2$, providing compelling evidence that the crossover from SR to LR universality occurs exactly at $\sigma = 2$, in contradiction to Sak's criterion.

For the Ising case ($q=2$), we perform a similar analysis using the FK representation, focusing on the Binder ratio $Q_m = \langle M^2 \rangle^2 / \langle M^4 \rangle $ and the FK critical polynomial $R_p$. Although an event-based method is unavailable for the Ising model, we carry out large-scale simulations up to $L = 8192$ to address finite-size corrections. Consistent with the percolation results, the geometric observable $R_p$ shows a pronounced deviation from zero and a sharp discontinuity at $\sigma = 2$, providing unambiguous evidence for a change of universality. In contrast, the Binder ratio $Q_m$ is hindered by significant statistical errors, and thus, the jump is not detected (see Fig.~\ref{fig:PD_Rp}). Furthermore, the critical exponent $\eta$ also exhibits a discontinuous jump at $\sigma=2$.  

Finally, it is worth noting the 2D LR-UF model. Similar to percolation, this is a purely graphical model but with a continuous symmetry~\cite{PhysRevLett.93.080601}. Despite its simplicity, the model exhibits rich critical behavior: for $\sigma > 2$, no finite-temperature phase transition occurs, while for $\sigma \le 2$, a second-order phase transition appears. The resulting phase diagram closely resembles that of the 2D LR-Heisenberg model.

In summary, all the 2D LR models, including Ising, XY, Heisenberg, percolation, and UF, consistently support a universality boundary at $\sigma_* = 2$. In the next section, we present a detailed numerical analysis of the 2D LR-Ising model, which has historically been at the center of this debate.

\section{Results for 2D Long-range Ising Model}
\label{sec:results}

\begin{table*}[!ht]
    \centering
    \caption{Summary of the critical points $\beta_c$, the universal values ($R_p$ and $Q_m$) and critical exponents ($\eta$ and $y_t$) near $\sigma = 2$ for 2D LR-Ising model, with `NN' denoting the nearest-neighbor case for reference. Previous results from the literature are also included for comparison.}
    \begin{tabular}{l|l|l|lll|lll|l}
    \hline\hline
        $~~~\sigma$ & $~~~~~~~\beta_c$ & ~~~$R_{p,c}$ & \multicolumn{3}{c|}{$Q_{m,c}$} & \multicolumn{3}{c|}{$\eta$} & $y_t~(1/\nu)$  \\ \hline
        1.75  & 0.329 136(1) & -0.342(5) & 0.79(1)  & ~~~0.84(1)\cite{Luijten2002} & 0.815(32)\cite{horita2017} & 0.335(4) & ~~~0.286(24)\cite{Luijten2002} & 0.332(8)\cite{picco2012} & 0.999(6)  \\ 
        1.875 & 0.336 985(2) & -0.207(9) & 0.815(8) &   &    & 0.293(3) &  &  & 1.008(9)  \\ 
        2     & 0.344 439(2) & -0.065(8) & 0.855(6) & ~~~0.850(6)\cite{Luijten2002} & 0.862(18)\cite{horita2017} & 0.273(3) & ~~~0.266(16)\cite{Luijten2002} & 0.262(4)\cite{picco2012} & 1.002(7)  \\ \hline
        2.2   & 0.355 388(2) & ~0.005(6) & 0.855(3) &  &  & -        &  &  & -        \\ 
        2.5   & 0.369 446(2) & ~0.001(2) & 0.857(1) & ~~~0.860(3)\cite{Luijten2002} &  & 0.250(1) & ~~~0.246(8)\cite{Luijten2002} &  & 1.007(5)   \\ 
        NN    & 0.440 686...\cite{PhysRev.65.117}& ~0~\cite{Jacobsen_2014} & \multicolumn{3}{l|}{0.856 216(1)\cite{Kamieniarz_1993} } & \multicolumn{3}{l|}{0.25~\cite{PhysRev.65.117}} & 1~\cite{PhysRev.65.117}  \\ \hline\hline
    \end{tabular}
    \label{tab:summary}
\end{table*}

In this section, we present our simulation results for the 2D LR-Ising model, which provide evidence that the universality boundary between the LR and SR regimes lies at $\sigma_* = 2$. To achieve this, we employ an enhanced variant of the Luijten–Blöte cluster algorithm~\cite{luijten1995, luijten1997, michel2019}, enabling simulations up to $L = 8192$. We extract the critical points and universal values from the critical polynomial $R_p$ and the Binder ratio $Q_m$, and obtain the critical exponents from the susceptibility $\chi$ and the scaled covariance $g^{(x)}_{ER}$ (defined later in Eq.~\eqref{eq:gx}). By comparing with previous works~\cite{Luijten2002,angelini2014,horita2017}, we find that our numerical results are more precise with smaller errors, clearly demonstrating the location of $\sigma_*=2$. Our results are summarized in Table~\ref{tab:summary}.

\subsection{Critical points}
\label{subsec:Tc}

The 2D LR-Ising model exhibits a second-order transition across both the LR and SR regimes. Therefore, the critical points $\beta_c = 1 / T_c$ can be determined by analyzing the crossing behaviors of dimensionless quantities. We measure the Binder ratio of magnetization, $Q_m$, and the critical polynomial, $R_p$, as dimensionless quantities to estimate the critical points (for their definitions, see Sec.~\ref{sec:review}).

In Fig.~\ref{fig:Overview} in Appendix~A, we show the overall crossing behavior of the two dimensionless quantities $R_p$ and $Q_m$ at $\sigma = 1.75$, $1.875$, and $2$. At first glance, clear crossing points can be observed, indicating the locations of the critical points from the horizontal coordinates of the crossings.

To systematically estimate the critical points and universal values, we perform least-squares fits of the MC data for the critical polynomial $R_p$ and the Binder ratio $Q_m$ using the standard FSS ansatz of second-order transitions:
\begin{align}
    \mathcal{O}(\beta, L) = \mathcal{O}_{c} + \sum_{k=1}^l a_k [(\beta_c - \beta) L^{y_t}]^{k} \nonumber \\ 
     + b_1 L^{-y_1} + b_2 L^{-y_2} + c_1 (\beta_c - \beta) L^{-y_1 + y_t},
     \label{eq:beta_c_fiting}
\end{align}
where $\mathcal{O} = R_p, Q_m$. Here $l$ is the highest order retained in the fitting ansatz, $y_t=1/\nu$ is the thermal scaling exponent, and $y_2 > y_1 > 0$ are the finite-size correction exponents. The last term accounts for the coupling between the correction and scaling variables. The fitting parameter $\beta_c$ represents the critical point, and $\mathcal{O}_{c}$ represents the universal value of $R_p$ or $Q_m$ at criticality.

The fitting details are provided in Appendix~A, and the fitting results are summarized in Table~\ref{tab:Rp} for $R_p$ and Table~\ref{tab:Qm} for $Q_m$. In these tables, both $R_p$ and $Q_m$ yield consistent critical points within the error bars.

Moreover, to further verify the reliability of the critical-point estimation, we analyze the finite-size pseudo-critical points $\beta_L$, defined as the horizontal coordinates of the intersections between the $R_p$ (or $Q_m$) curves for two consecutive system sizes, $L/2$ and $L$. As $L$ increases, $\beta_L$ converges to the true critical inverse temperature $\beta_c$ in the thermodynamic limit.

\begin{figure*}[!ht]
    \centering
    \includegraphics[width=\linewidth]{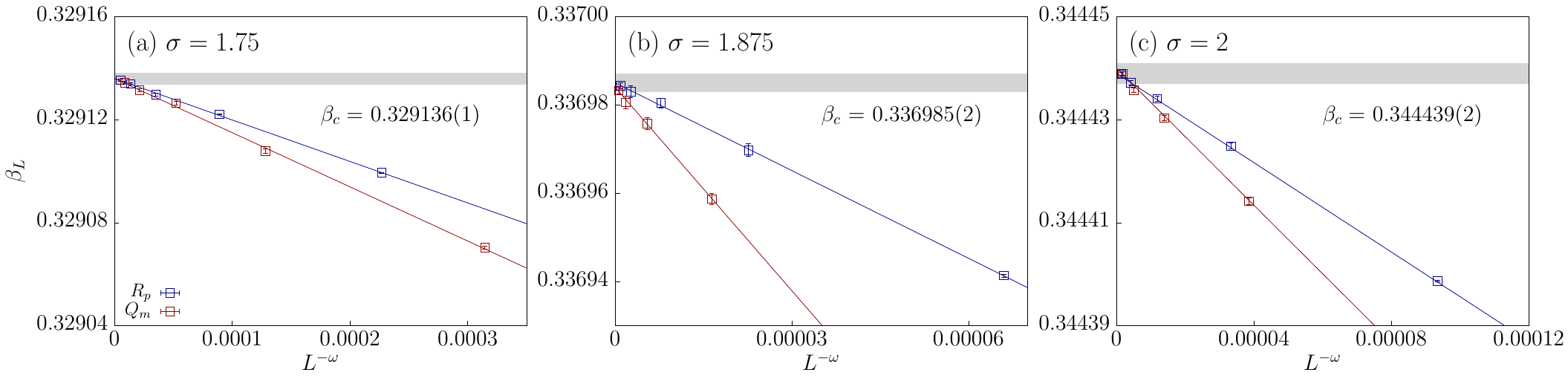}
    \caption{Extrapolation of the crossing points $\beta_L$ for $R_p$ (blue) and $Q_m$ (red) at different $\sigma$. Both quantities yield consistent estimates of the critical inverse temperature $\beta_c$ for $\sigma = 1.75$ (a), $1.875$ (b), and $2$ (c). The horizontal axis represents the correction term $L^{-\omega}$, with $\omega = 1.34$ for $R_p$ and $1.29$ for $Q_m$ in panel (a), $\omega = 1.54$ and $1.59$ in (b), and $\omega = 1.48$ and $1.46$ in (c). The shaded bands indicate the extrapolated $\beta_c$ as $L \to \infty$.}
    \label{fig:Kc_extra}
\end{figure*}

To quantify this convergence, we perform least-squares fits of $\beta_L$ to the standard FSS form
\begin{align}
    \beta_L = \beta_c + a L^{-\omega},
    \label{eq:beta_c_extrap}
\end{align}
where $a$ is a non-universal amplitude and $\omega = y_t + y_1$, with $y_t$ and $y_1$ denoting the thermal and leading irrelevant exponents, respectively.  
This relation can be derived from the FSS expansion in Eq.~(\ref{eq:beta_c_fiting}): by considering the intersection condition between two system sizes and retaining the leading correction term $L^{-y_1}$, one finds that the pseudo-critical shift scales as $L^{-y_t - y_1}$, hence $\omega = y_t + y_1$.

Since the extrapolation form includes only the leading correction term $a L^{-\omega}$, we use data from relatively large system sizes to suppress subleading effects whenever possible. We find that using data with $L \ge 256$ yields reliable results. A similar consideration also applies to the FSS fitting in Eq.~(\ref{eq:beta_c_fiting}), where setting $L_{\min} = 256$ effectively eliminates small-size corrections and leads to stable fits (see Appendix~A for details).

As shown in Fig.~\ref{fig:Kc_extra}, the extrapolated $\beta_c$ values obtained from both $R_p$ and $Q_m$ converge to a common limit within uncertainty, confirming the consistency between the two independent observables.
The $\omega$ values used in the extrapolation are listed in the caption of Fig.~\ref{fig:Kc_extra}.
The fitted $\beta_c$ values from Eq.~(\ref{eq:beta_c_fiting}) fall within the extrapolation error bars, further confirming the reliability of our analysis. Therefore, for conciseness, we only list the extrapolated results in Table~\ref{tab:summary}.

It is worth noting that Horita \textit{et al.}~\cite{horita2017} also adopted a normalized formulation for the LR-Ising model, allowing a direct comparison of the estimated critical points. After converting their estimations to our normalization conventions, the critical points are fully consistent: Horita \textit{et al.}~\cite{horita2017} reported $\beta_c = 0.329\,134(8)$ for $\sigma = 1.75$ and $\beta_c = 0.344\,445(11)$ for $\sigma = 2$, while our results are $\beta_c = 0.329\,136(1)$ and $\beta_c = 0.344\,439(2)$, respectively. These values agree within the error bars, with our simulations offering significantly higher precision.

For completeness, we have also determined the critical points in the SR regime, including $\sigma = 2.2$, $2.5$, and the NN limit. The analysis follows the same procedure and will not be repeated here. Note that the NN case allows for an exact calculation of $R_p$, and $R_p$ shows no finite-size corrections, resulting in a beautiful plot illustrated in Fig.~\ref{fig:Rp_SR}(a) in Appendix~A.

Note that the other critical points shown in Fig.~\ref{fig:PD}(a) are determined with much less precision, as they are mainly used for drawing the phase diagram; therefore, the detailed results are not shown here.

\subsection{Dimensionless quantities}
\label{subsec:Rp}

\begin{figure*}[!ht]
    \centering
    \includegraphics[width=\linewidth]{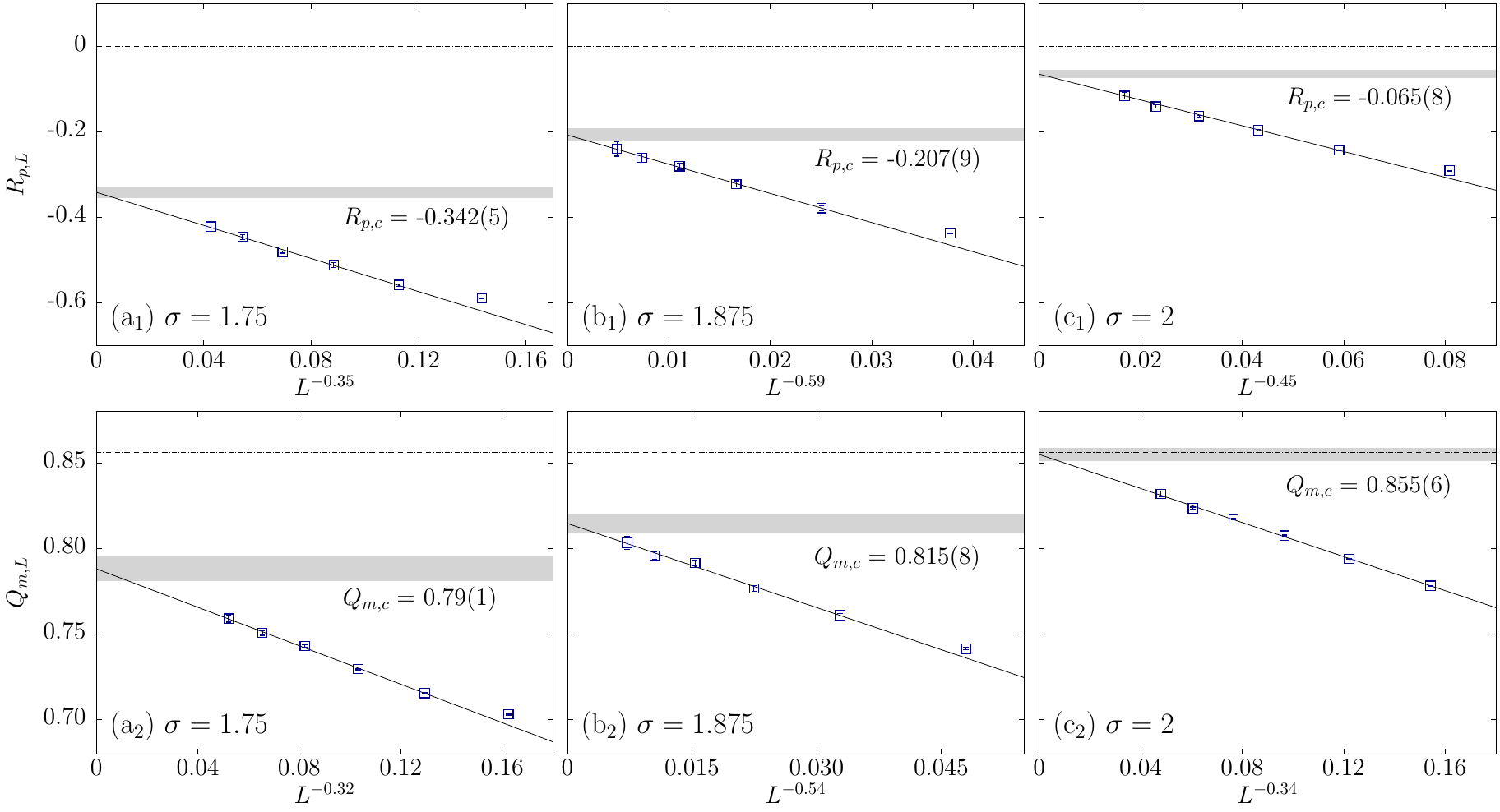}
    \caption{Extrapolation of the crossing points of $R_p$ and $Q_m$ for different system sizes. The data points represent the crossing value $\mathcal{O}_L$ between $(L/2, L)$, fitted using Eq.~(\ref{eq:universal_extrap}). Deviations from the SR universal values, 0 for the critical polynomial $R_p$ and 0.856216(1) for the Binder ratio $Q_m$, are observed for $\sigma \leq 2$. The gray bands indicate the converged values obtained from extrapolation.}
    \label{fig:Qm_LR}
\end{figure*}

For second-order transitions, the universal value of a dimensionless quantity is defined as its value at criticality, which reflects the universality class of the system. For the 2D LR-Ising model, it remains at the NN value for $\sigma > \sigma_*$, drifts continuously in the regime $1 < \sigma < \sigma_*$, and becomes constant again for $\sigma < 1$. Most previous analyses relied on the Binder ratio $Q_m$, but its universal value varies only slightly near $\sigma \in [1.75, 2]$ and is affected by significant uncertainties, making it difficult to identify subtle changes in universality. Here, we instead adopt a geometric perspective and employ the FK critical polynomial $R_p$, which exhibits significantly smaller finite-size corrections and a strong contrast between distinct universality classes. For completeness, we also measured the Binder ratio $Q_m$ for comparison.

As shown in Fig.~\ref{fig:Overview} in Appendix~A, one can already see with significant confidence that in the limit of large system size $L \to \infty$, the universal values for $\sigma = 1.75$ and $1.875$ deviate from those of the SR case. Specifically, the universal value of $R_p$ shifts significantly away from zero~\cite{Jacobsen_2014}, while $Q_m$ clearly departs from $0.856216(1)$~\cite{blote1995}. At $\sigma = 2$, a noticeable deviation is still observed for $R_p$, whereas for $Q_m$, the difference remains within the error bars and is therefore hardly distinguishable.

From the analysis of FSS least-squares fits via Eq.~\eqref{eq:beta_c_fiting}, the 
quantification of these deviations is given in Table~\ref{tab:Rp} for $R_p$ and Table~\ref{tab:Qm} for $Q_m$. 
Although the fitted results vary slightly with different correction exponents, we can clearly see that for $\sigma = 1.75$ and $1.875$, they deviate from the short-range values. 
For $\sigma = 2$, only $R_p$ shows a noticeable deviation from zero, which is consistent with our expectations.

Additionally, similar to the analysis of $\beta_L$, we can provide a more intuitive extrapolation for $\mathcal{O}_L$, defined as the vertical coordinates of the crossing points between curves for system sizes $L/2$ and $L$. These values converge to the thermodynamic universal value $\mathcal{O}_c$ as the system size $L$ increases. The $L$-dependent FSS form follows
\begin{align}
    \mathcal{O}_{L} = \mathcal{O}_{c} + a L^{-\omega},
    \label{eq:universal_extrap}
\end{align}
where $\mathcal{O} = R_p$ or $Q_m$, and $a$ is a non-universal amplitude. Here, $\omega = y_1$ represents the contribution from irrelevant fields.  
This relation can be derived from Eq.~\eqref{eq:beta_c_fiting} by setting $\beta = \beta_c$ and keeping only the correction term $L^{-y_1}$.  
It allows a straightforward linear extrapolation in $L^{-\omega}$ to extract $\mathcal{O}_c$.  
Similar to the extraction of the critical points, we use data with $L \ge 256$ to obtain reliable results.

Figure~\ref{fig:Qm_LR} shows the extrapolation results for both observables. The extrapolated results, highlighted by shaded bands, confirm the bare-eye observations of Fig.~\ref{fig:Overview} and the fitting results by Eq.~\eqref{eq:beta_c_fiting}: $R_p$ deviates substantially from zero for $\sigma \le 2$, signifying a clear shift in universality. Notably, at $\sigma = 2$, a visible gap appears between the extrapolated $R_p$ and the NN value, implying that the system at $\sigma = 2$ has already departed from the SR universality class. For $Q_m$, the deviation from $Q_{\rm SR}=0.856216(1)$ is clearly observed at $\sigma = 1.75$ and $1.875$, but for $\sigma = 2$ it is much smaller and largely buried within statistical uncertainties.

It is interesting that the $\omega$ values in Eq.~\eqref{eq:beta_c_extrap} and Eq.~\eqref{eq:universal_extrap} correspond to $y_t + y_1$ and $y_1$, respectively. Note that in Table~\ref{tab:Rp} and Table~\ref{tab:Qm}, we find that $y_t \approx 1$ at $\sigma = 1.75, 1.875,$ and $2$. Therefore, if we subtract one from the $\omega$ values in Fig.~\ref{fig:Kc_extra}, the resulting values are very close to those of $\omega$ in Fig.~\ref{fig:Qm_LR}, which further supports the self-consistency of our extrapolation.

\begin{figure}[t]
    \centering
    \includegraphics[width=\linewidth]{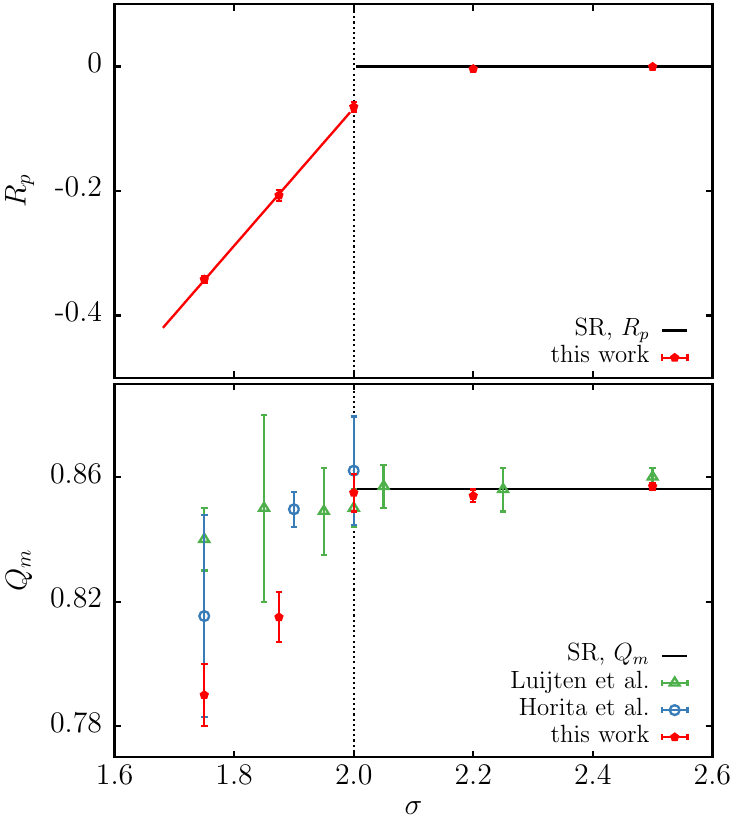}
    \caption{Extrapolated universal values of $R_p$ and $Q_m$ as functions of $\sigma$. The black lines denote the SR universal values ($R_p = 0$, $Q_m = 0.856216$). A clear jump is observed at $\sigma = 2$ for $R_p$, marking the crossover from SR to LR universality. For $\sigma < 2$, both quantities continuously decrease from the SR values, while for $\sigma > 2$, they remain consistent with the SR values. Previous studies~\cite{Luijten2002, horita2017} are also shown for comparison.}
    \label{fig:PD_Rp}
\end{figure}

To provide an overall perspective, we summarize in Fig.~\ref{fig:PD_Rp} the extrapolated critical values of $R_p$ and $Q_m$ as functions of $\sigma$. For $Q_m$, we also include a comparison with previous studies~\cite{Luijten2002, horita2017}. As $\sigma$ decreases from the SR to the LR regime, $R_p$ shows a clear discontinuity at $\sigma = 2$, followed by a gradual decline for $\sigma < 2$. This discontinuity is particularly pronounced in $R_p$, whereas for $Q_m$ it is less evident due to larger statistical uncertainties. Compared with previous results, our extrapolated values of $Q_m$ fall well within their reported error bars. However, while our results clearly reveal a universality change as $\sigma$ decreases from $2$, the previous studies could hardly distinguish such a transition because of the large uncertainties. The extrapolated universal values for $R_p$ and $Q_m$ are summarized in Table~\ref{tab:summary}.

Finally, we remark on the approach of Ref.~\cite{horita2017}, which introduced a nonlinear ``self-combined'' Binder ratio,
\begin{equation}
S(Q_m) = \frac{Q_m}{Q_{\rm SR}} + \frac{Q_{\rm SR}}{Q_m} - 2,
\end{equation}
with $Q_{\rm SR} = 0.856216$. The motivation of this construction was to define a dimensionless quantity that suppresses FSS corrections, based on the assumption that the deviation of $Q_m$ from its SR value for $\sigma \in [1.75, 2]$ originates mainly from finite-size effects rather than from a genuine change of universality. This quantity indeed makes the observable almost insensitive to FSS corrections, as reflected in their results where $S(Q_m)$ approaches zero within $\sigma \in [1.75, 2]$, seemingly supporting Sak's criterion. However, such nonlinear manipulation inevitably compresses genuine deviations between distinct universality classes and may obscure real physical differences near $\sigma = 2$. For instance, when $Q_m = 0.8$, its deviation from the SR value $Q_{\rm SR}$ is about $0.056$, yet substituting it into $S(Q_m)$ yields a deviation of only about $0.0046$, smaller by an order of magnitude. As a result, $S(Q_m)$ becomes difficult to use for locating critical points or identifying universality changes. In contrast, our geometric observable $R_p$ provides a direct, model-independent, and physically transparent probe of universality The sharp jump in $R_p$ at $\sigma = 2$ therefore provides compelling evidence that the 2D LR-Ising model undergoes a genuine change of universality at this point, in contradiction to Sak's criterion.

\subsection{Critical exponents}
\label{subsec:eta}

Critical exponents provide a direct characterization of universality, and consequently, much of the existing literature on this issue has focused on their estimation~\cite{Luijten2002, picco2012, angelini2014}.
However, obtaining reliable exponents is challenging because the precise location of the critical point is often uncertain, and strong finite-size effects can obscure the true asymptotic behavior. 
Here, based on critical points determined with high precision above and simulations of large system sizes up to $L = 8192$, we achieve a more accurate extraction of the exponents. Moreover, the availability of numerous previous studies enables a direct and meaningful comparison of our results with earlier findings.

\begin{figure}
    \centering
    \includegraphics[width=0.9\linewidth]{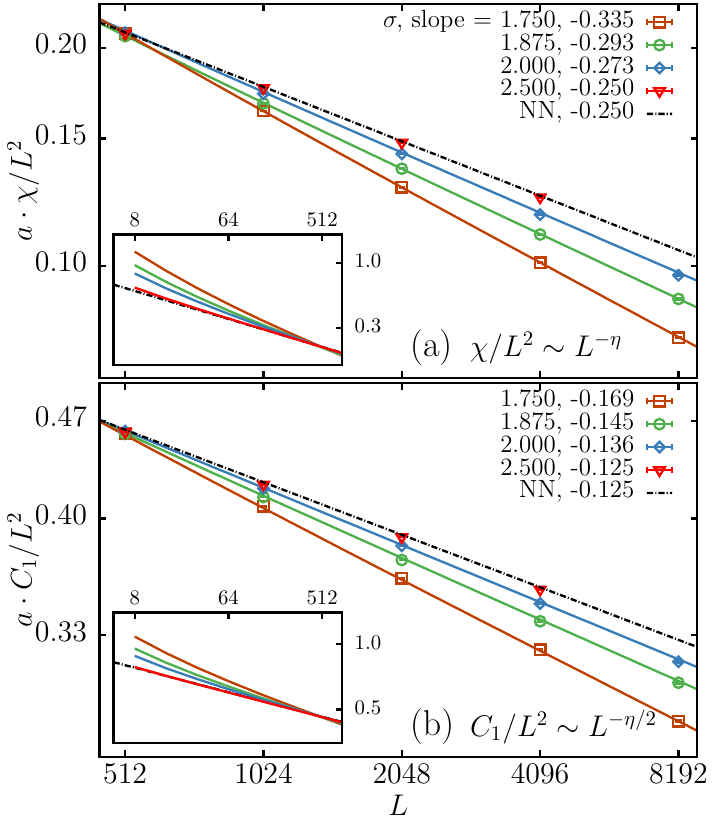}
    \caption{Critical exponent $\eta$ of the 2D LR-Ising model for $\sigma = 1.75$, $1.875$, $2$, $2.5$, and the NN case. (a) According to the scaling relation $\chi / L^2 \sim L^{-\eta}$, the slopes of the log-log plot indicate that $\eta$ starts to deviate from $0.25$ for $\sigma \leq 2$. (b) According to the scaling $C_1 /L^2 \sim L^{-\eta/2}$, the slopes of the log-log plot similarly suggest a deviation from $\eta = 0.125$ for $\sigma \leq 2$. 
    The parameter $a$ is used to vertically shift the curves for visual clarity. The inset shows the data for system sizes from $L=8$ to 512, which suffer significantly from finite-size corrections. In contrast, the data points for large system sizes follow rather straight lines in the log-log plot, indicating the reliability of the extracted critical exponent $\eta$.
}
    \label{fig:eta_scaling}
\end{figure}

Firstly, we focus on the anomalous dimension $\eta$. At criticality, the finite-size scaling of the susceptibility satisfies $\chi \sim L^{2-\eta}$; therefore, we plot $\chi/L^2$ versus $L$ on a log-log scale in Fig.~\ref{fig:eta_scaling}(a), where the slope is $-\eta$. Due to the strong finite-size corrections induced by LR interactions, the slope gradually converges only for $L \ge 512$. In Fig.~\ref{fig:eta_scaling}(a), where the data are rescaled by a constant $a$ such that the $L=512$ data for different $\sigma$s are approximately the same, one can clearly observe that the slopes for $\sigma = 1.75$, $1.875$, and $2$ differ entirely from those for $\sigma = 2.5$ and the nearest-neighbor case. Further, as $\sigma$ decreases, the value of $\eta$ clearly increases. This indicates that $\eta$ at $\sigma = 2$ has already departed from the SR value.

Moreover, from the perspective of the geometric FK representation, the fraction of the largest-cluster size $C_1$ over the whole lattice can also act as an order parameter, similar to the magnetization density. Namely, the size of the largest cluster scales as $C_1 \sim L^{2 - \eta/2}$ in 2D. We plot $C_1/L^2$ versus $L$ on a log-log scale, and similarly, it is clearly seen that when $\sigma \ge 2$, the slope $-\eta/2$ deviates from the SR case.
In short, with sufficiently large system sizes in a wide range from $L = 512$ to $L=8192$, the bare-eye view of the finite-size scaling of $\chi$ and $C_1$ can already yield strong evidence that $\sigma_*=2$. In contrast,  the data for $L \leq 512$ suffer from strong finite-size corrections (see insets of Fig.~\ref{fig:eta_scaling}), and, to extract correct values of $\eta$, these corrections should be carefully and systematically taken into account. Note that the maximum system sizes are $L=1000$ in Ref~\cite{Luijten2002} and $5120$ in Ref~\cite{picco2012}.

To quantitatively extract $\eta$, we fit $\chi$ and $C_1$ to the finite-size scaling form given in Eq.~\eqref{eq:scaling_fitting}, with the fitting details provided in Appendix~B. The results are shown in Fig.~\ref{fig:eta_scaling}, where the straight lines guiding the data points correspond to the leading terms of the fit results, i.e., $L^{-\eta}$ for Fig.~\ref{fig:eta_scaling}(a) and $L^{-\eta/2}$ for Fig.~\ref{fig:eta_scaling}(b). We find that both estimators yield consistent estimates of $\eta$.

\begin{figure}
    \centering
    \includegraphics[width=0.9\linewidth]{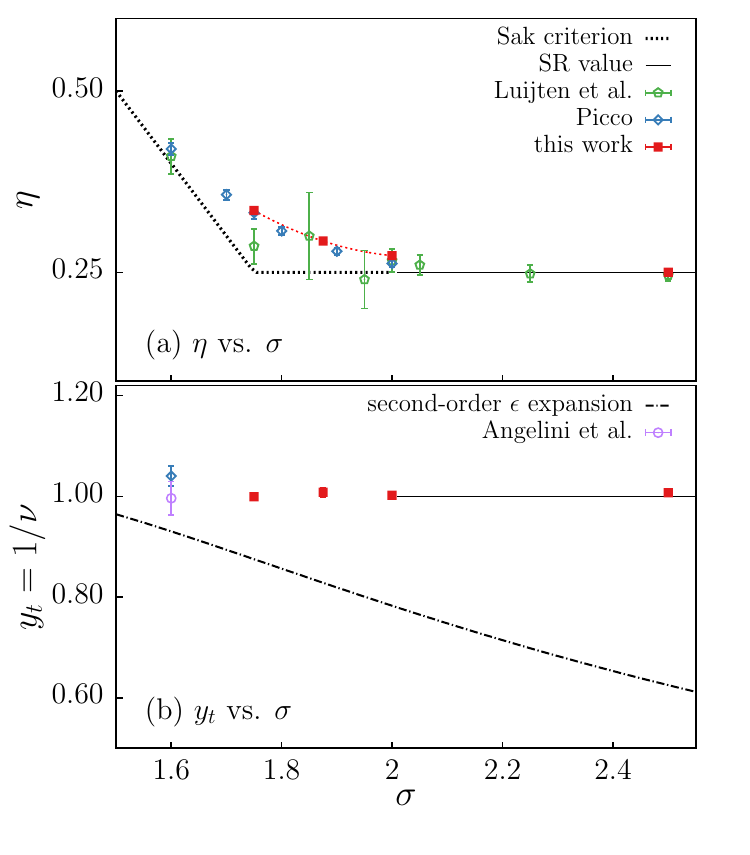}
    \caption{The value of critical exponents ($\eta$ and $y_t$) at various $\sigma$. The solid lines denote the SR values for $\eta$ and $y_t$. (a) The red dots denote the estimates of $\eta$ in this work. Luijten and Picco's results are also contained~\cite{Luijten2002, picco2012}, respectively denoted by green and blue dots. The dotted lines correspond to Sak's prediction: $\eta = \text{max}(2-\eta,\frac{1}{4})$. Obviously, our results deviate from Sak's prediction, suggesting that for $1.75 < \sigma \leq 2$, there exists a discontinuous jump at $\sigma = 2$, indicating that the short-range behavior is not recovered. (b) The red dots denote the estimates of $y_t$ in this work. The dotted line represents the results from second-order $\epsilon$ expansion~\cite{fisher1972} (note that $y_t = 1/\nu$).}
    \label{fig:eta_yt}
\end{figure}

In Fig.~\ref{fig:eta_yt}(a), we plot our estimates of $\eta$ as a function of $\sigma$. According to Sak's criterion, $\eta = \max (2 - \sigma, \eta_{\rm SR})$, where $\eta_{\rm SR} = 1/4$ for the 2D Ising model, as shown by the dashed line. Our estimates of $\eta$ are plotted as red dots. It is clear that there is a deviation from Sak's criterion for $\sigma \in [1.75,2]$. Notably, at $\sigma = 2$, $\eta$ is already larger than $1/4$, which is consistent with the universal value of $R_p$ discussed in the previous section. Moreover, other previous numerical results are also plotted here. The green dots from Luijten et al.~\cite{Luijten2002}, which have large errors in this regime, cannot clearly distinguish the deviation; specifically, their central values are actually close to our results, but the errors are very large. As for the blue dots from Picco~\cite{picco2012}, they are very close to our results, and his estimates, extending to $\sigma = 1.6$, also imply that the deviation already starts near $\sigma = 1.6$.

Then, for the thermal exponent $y_t \equiv 1/\nu$, we use the scaled covariance between the wrapping indicator and the NN energy,  
\begin{align}
    g^{(x)}_{ER} \;=\; \langle \mathcal{R}^{(x)}\rangle\langle \mathcal{E}\rangle - \langle \mathcal{R}^{(x)} \mathcal{E} \rangle,
    \label{eq:gx}
\end{align}
where $\mathcal{E} = -J_{\rm nn}\sum_{\langle i,j\rangle}S_i S_j$ denotes the NN energy density, and $J_{\rm nn} = c(\sigma, L)$ is the NN coupling strength. Here, $\mathcal{R}^{(x)} = 1$ if at least one cluster wraps in the $x$ direction (otherwise $\mathcal{R}^{(x)} = 0$). This covariance scales as $g^{(x)}_{ER} \sim L^{y_t}$~\cite{PhysRevE.99.042150}, which can be used to extract the exponent $y_t$.

The fits of $g^{(x)}_{ER}$ yield $y_t$ values that are very close to $1$ for all investigated $\sigma \leq 2$ (the NN case has $y_t = 1$ exactly), and the fitting details are provided in Appendix~B. It is difficult to use $y_t$, like $\eta$, as an indicator of a change in universality to distinguish between $\sigma \geq 2$ and $\sigma > 2$.


In Fig.~\ref{fig:eta_yt}(b), we plot the estimates of $y_t$ as functions of $\sigma$. The blue and purple dots at $\sigma = 1.6$ correspond to the results from Picco~\cite{picco2012} and Angelini et al.~\cite{angelini2014}, respectively. The dashed line represents the second-order $\epsilon$ expansion from Fisher et al.~\cite{fisher1972}. We can see that $y_t$ remains close to $1$ across the investigated range, making it less effective as an indicator of universality. The deviations from the $\epsilon$ expansion are also quite clear, which again imply the limitation of the mean-field expansion, similar to the case of $\eta$. 

Finally, to demonstrate the reliability of the estimated critical exponents $\eta$ and $1/\nu$ (including their central values and quoted error bars in Table I), we plot Fig~\ref{fig:eta_precise} and Fig~\ref{fig:yt_precise}.

The estimates of the two exponents, $\eta$ and $y_t$, together with comparisons to previous results, are summarized in Table~\ref{tab:summary}.


\section{Scenario for the crossover from short-range to long-range universality}
\label{sec:scenario}

Sak's criterion $\sigma_* = 2 - \eta_{\rm SR}$ has long been regarded as a natural refinement of the original Fisher picture, ensuring a continuous function of $\eta(\sigma)$ across the LR–SR boundary~\cite{sak1973}. 
Given the formidable complexity of field theories with LR interactions, Sak’s formulation represented a substantial theoretical achievement. Nevertheless, upon closer examination, one finds that the argument ultimately rests on a set of assumptions whose validity may be limited.

First, the relation $\eta = 2 - \sigma$ originates from perturbative RG expansions~\cite{fisher1972} around the Gaussian fixed point, i.e., near $\sigma = d/2$,
\begin{align}
\eta = 2 - \sigma + O(\epsilon^3), \quad \frac{1}{\nu} = \sigma - \frac{(n+2)}{(n+8)} \epsilon + O(\epsilon^2),
\label{eq:eta_expend}
\end{align}
with $\epsilon = 2\sigma - d \geq 0$. This expansion is obtained by fixing $\sigma$ and expanding in $d$. In the limit $n \to \infty$, it recovers the LR spherical model~\cite{PhysRev.146.349} result $1/\nu = d - \sigma$.  
However, if one fixes $d$ and expands in $\sigma$, the expression becomes
\begin{align}
\nu = \frac{1}{\sigma} + \frac{4}{\sigma d} \frac{(n+2)}{(n+8)} \Delta \sigma + O(\Delta \sigma^2),
\end{align}
with $\Delta \sigma = \sigma - \frac{1}{2}d \geq 0$. This expression does not recover the LR spherical model result in the limit $n \to \infty$, revealing the subtlety of the LR perturbative expansion.

Moreover, although this form remains surprisingly accurate up to three loops for $\eta$, there is no solid theoretical reason for it to hold as $\sigma \to 2$.
Actually, in Eq.~\eqref{eq:eta_expend}, the correlation-length exponent $\nu$ shows a clear dependence on parameters $\epsilon$, $n$, and $d$, indicating that the universality class cannot be fully characterized by $\eta$ alone. 
It remains an open question whether the perturbative result remains valid for a wide range from $\sigma=d/2$ to $2 - \eta_{\rm SR}$, an assumption underlying Sak's argument.

Second, discontinuous changes of critical exponents in parameter space are not pathological but rather expected within the RG framework. Transitions from second-order to first-order, or jumps in exponents at tricritical or multicritical points, are well-established phenomena~\cite{blumeTheoryFirstOrderMagnetic1966,capelPossibilityFirstorderPhase1966,blumeIsingModelTransition1971}. Therefore, a discontinuity of $\eta$ at $\sigma_*$ cannot by itself be used to dismiss the scenario $\sigma_* = 2$.

Third, several physical inconsistencies arise when applying $\sigma_* = 2 - \eta_{\rm SR}$ across different models:
\begin{itemize}
    \item For percolation in $ 2 <d <6$, one has $\eta_{\rm SR} <0$, leading to $\sigma_* > 2$, which is physically unlikely.
    \item For the 2D XY model, $\eta_{\rm SR}(T, \sigma)$ depends continuously on both $T$ and $\sigma$ for $\sigma > 2$, rendering the boundary $\sigma_* = 2 - \eta_{\rm SR}$ ill-defined.
    \item For the 2D Heisenberg model, which lacks a finite-temperature transition, $\eta_{\rm SR}$ is ill-defined.
\end{itemize}
It stems from the continuity assumption of $\eta$ and the perturbative relation $\eta = 2 - \sigma$, yet neither of these premises is robust. Hence, $\sigma_* = 2 - \eta_{\rm SR}$ cannot serve as a universal criterion.

Fourth, the Sak's criterion predicts that the SR universality should extend from $\sigma > 2$ to  $\sigma> 2-\eta_{\rm SR}$. The ``bare" (before being renormalized) field theoretical description of the O($n$) spin systems can be approximately written as 
\begin{align}
    \beta H=&\int{\frac{\mathrm d^dq}{(2\pi)^d}(\frac t2+\frac {K_2}{2} q^2+K_\sigma q^\sigma)\boldsymbol{\Psi}(\boldsymbol{q})\cdot\boldsymbol{\Psi}(-\boldsymbol{q})} 
    \nonumber \\
    & + \int \mathrm{d}^d r |\boldsymbol{\Psi}(\boldsymbol{r})|^4 \; ,
\label{eq:lr-phi4}
\end{align}
where $\boldsymbol{\Psi}(\boldsymbol{q})$ is the field in the momentum space.
It can be seen that the non-analytical $q^\sigma$ term dominates over the regular $q^2$ kinetic term in the long-wave (small-$q$) limit, which is crucial for critical phenomena. 
An argument for the Sak's criterion is then that, after renormalization, the dominant term $q^\sigma$ becomes irrelevant, and the subleading term $q^2$ becomes dominant. However, to our knowledge, no solid RG calculations around  $\sigma=2$ are available to support this unusual argument.

In contrast, the dominance of the $q^\sigma$ term over the $q^2$ term can be demonstrated in the Goldstone-mode physics of the LR-XY and LR-Heisenberg models in the low-T LRO phase, and can be proven for the LR-SRW (Le\'vy flight)~\cite{yao2024nonclassical,2dlrheisen}. 
For the LR-percolation and LR-FK-Ising model, this dominance can also be illustrated by the geometric structures of FK-random clusters at low temperatures. 
The crossover occurs where the analytic $q^2$ term overtakes the nonanalytic $q^{\sigma}$ contribution—namely at $\sigma_* = 2$.

Taken together, these considerations demonstrate that Sak's criterion, though elegant, rests on fragile assumptions. In contrast, a unified picture emerges from both theoretical reasoning and extensive numerical results across 2D systems -- XY, Heisenberg, percolation, and Ising~\cite{yao2024nonclassical,2dlrheisen,2dlrperco}. In all cases, the universality class changes sharply at $\sigma = 2$, where the anomalous dimension $\eta$ exhibits a weak but distinct jump. This establishes a fourth, self-consistent scenario: the exponents vary smoothly within the LR regime, yet the crossover at $\sigma_* = 2$ is discontinuous, indicating that the boundary point already belongs to the LR universality class rather than the SR one.

A more natural argument is that, for $\sigma>d/2$, the Gaussian fixed point becomes unstable, and the interplay of the non-analytic term $q^\sigma$, short-range fluctuations $q^2$, and interaction terms would lead to rich nonclassical critical behaviors. 
Based on insights from the LR-SRW (i.e., L\'evy flights), recent high-precision Monte Carlo results, and rigorous mathematical analyses, we propose three universality diagrams of the phase transitions in the $(d, \sigma)$ plane, respectively, for the percolation, O($n$) spin, and FK-Ising models~\cite{hdlr}. For the O$(n)$ spin models, the nonclassical regime occurs in the range $d/2 < \sigma \leq 2$, similarly to Ref.~\cite{fisher1972}.
The dynamics of the LR-SRW changes, from diffusive for $\sigma>2$, into superdiffusive for $\sigma \leq 2$ (logarithmically super-diffusive at $\sigma=2$).
Depending on $d$, this regime could be further divided into two sub-regimes separated by a new threshold $\sigma=1$.
For $\sigma<1$, the dynamics of the LR-SRW are hyper-ballistic, as dominated by extremely long-range jumps that effectively short-circuit the local lattice metric. 
It is conjectured~\cite{hdlr} that, in sub-regime $1 < \sigma \leq 2$, both $\eta$ and $\nu$ take non-trivial values (neither mean-field-like nor short-range-like), and, in sub-regime $d/2 < \sigma \leq 1$, the $\nu$ value is non-trivial but $\eta=2-\sigma$ holds exactly true. 
For $d \geq 2$, this is consistent with the fourth scenario in Fig.~\ref{fig:our_scenario}.


\section{Discussion}
\label{sec:discussion}

The present study completes a systematic investigation of long-range interacting systems in two dimensions by combining our previous results for the XY, Heisenberg, and percolation models with new large-scale simulations of the long-range Ising model. Across these models, which differ in their symmetries and in the nature of their phase transitions, our numerical analyses reveal a consistent and robust pattern: the crossover from long-range to short-range critical behavior takes place at $\sigma_* = 2$.

In the XY and Heisenberg models, the long-range interaction intrinsically changes the type of phase transition, providing a clear indication that the short-range description ceases to apply once $\sigma \le 2$. For long-range percolation, where the transition remains second order for all $\sigma$, several geometric observables, such as the critical polynomial $R_p$, can indicate the universality more sensitively, and we observe an abrupt change at $\sigma = 2$. These findings motivate a careful examination of the Ising model, whose transitions remain second order throughout and therefore require critical properties, rather than changes in the transition type, to identify the crossover.

By analyzing the critical polynomial $R_p$ in the Fortuin--Kasteleyn representation of 2D LR-Ising model together with the Binder ratio $Q_m$ and the anomalous dimension $\eta$, our simulations up to $L=8192$ provide convergent and mutually consistent evidence that the Ising universality class departs from the short-range behavior at $\sigma = 2$. The fact that three independent observables, both magnetic and geometric, lead to the same conclusion strongly suggests that $\sigma = 2$ marks the point at which the short-range fixed point loses stability. Compared with earlier numerical studies, the improved precision of our critical points and universal quantities allows this change to be resolved without ambiguity.

These numerical results also motivate a re-examination of the robustness of Sak's criterion. The perturbative assumption $\eta = 2 - \sigma$ is controlled only in the vicinity of $\sigma = d/2$ and becomes unreliable as $\sigma \to 2$. Moreover, continuity of critical exponents across the crossover is not guaranteed from the renormalization-group perspective; finite but abrupt changes can naturally arise when the controlling fixed point switches. A momentum-space analysis supports this viewpoint and indicates that the competition between the analytic $q^2$ term and the nonanalytic $q^{\sigma}$ term naturally identifies $\sigma = 2$ as the threshold separating long-range and short-range behavior. Although these considerations provide a coherent physical interpretation of our numerical findings, a fully controlled theoretical derivation in two dimensions remains open.

A deeper theoretical understanding of the regime near $\sigma = 2$, or possible experimental realizations in tunable long-range systems, remains an important direction for future work. More broadly, this problem constitutes a fundamental question in the physics of long-range interactions, providing a solid starting point for exploring long-range quantum systems and a basis for studying dynamical criticality in nonlocal models. It also calls for a renewed field-theoretical formulation of LR–O($n$) systems and associated renormalization-group analyses, which may eventually lead to a unified theoretical framework for long-range critical behavior.

\section*{Acknowledgements}
We thank Dingyun Yao for many stimulating discussions. The research is supported by the National Natural Science Foundation of China (NSFC) under Grant No. 12204173 and No. 12275263, as well as the Innovation Program for Quantum Science and Technology (under Grant No. 2021ZD0301900). YD is also supported by the Natural Science Foundation of Fujian Province 802 of China (Grant No. 2023J02032).


\appendix
\renewcommand{\thefigure}{A\arabic{figure}}
\renewcommand{\thetable}{A\arabic{table}}
\setcounter{figure}{0}
\setcounter{table}{0}

\renewcommand{\theHequation}{A\arabic{equation}}
\renewcommand{\theHfigure}{A\arabic{figure}}
\renewcommand{\theHtable}{A\arabic{table}}

\section{Least-squares fits of the critical points and universal values}

\begin{figure*}[t]
    \centering
    \includegraphics[width=\linewidth]{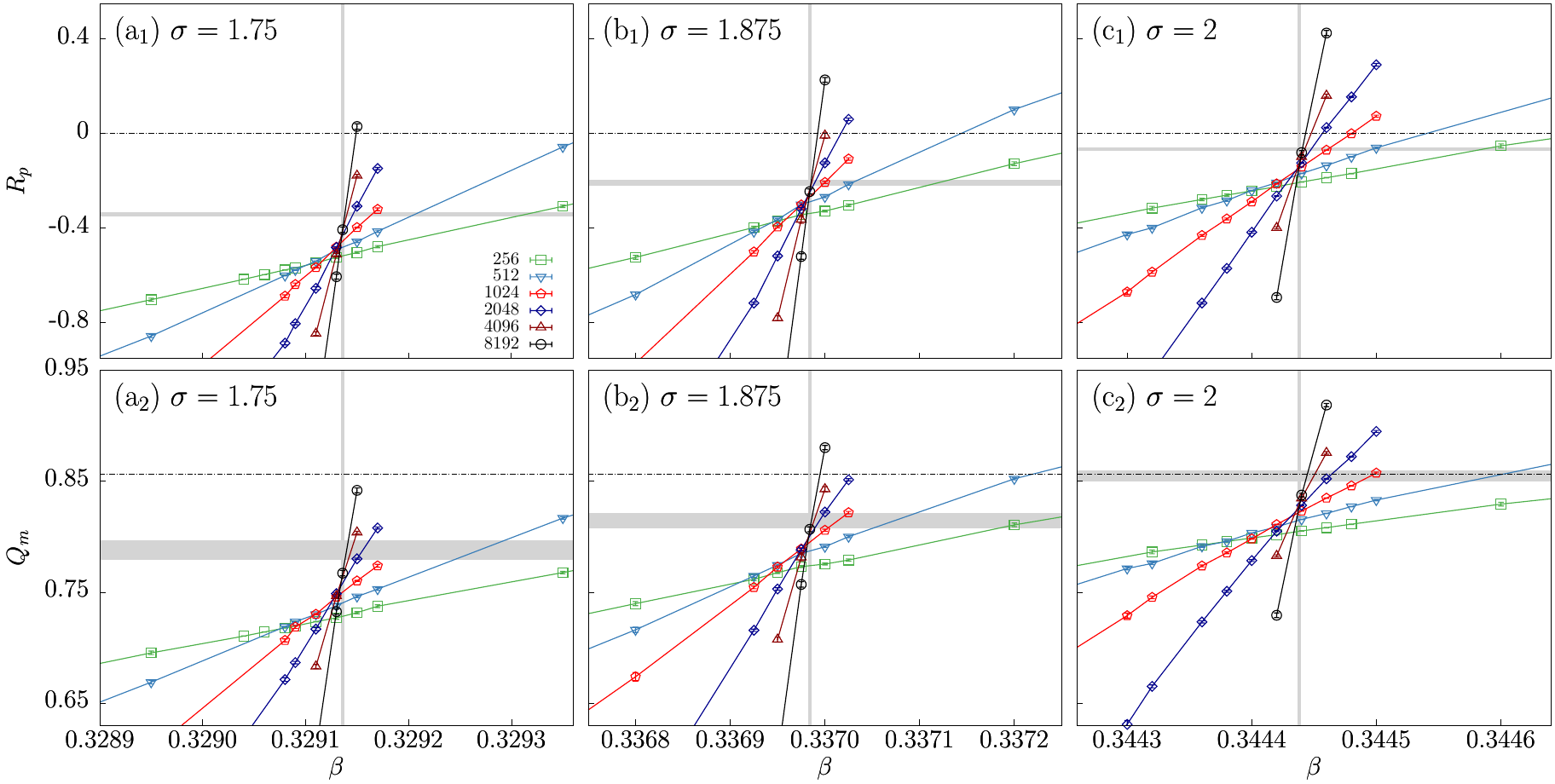}
    \caption{Overview of the dimensionless quantities $R_p$ and $Q_m$ for $\sigma = 1.75, 1.875, 2$, up to system size $L = 8192$. A clear crossing behavior can be observed, reflecting the location of criticality and universality. The dashed horizontal line denotes the SR universal values: 0 for $R_p$ and 0.856216 for $Q_m$. The deviation from the SR universal value at criticality is clearly visible.
    In panels ($\rm a_1$), ($\rm b_1$), and ($\rm c_1$), as the system size increases up to $L = 1024$, the universal value of $R_p$ gradually converges. The deviation from 0 is particularly pronounced for $\sigma = 1.75$, where it reaches nearly -0.4, a notably large value. It is also worth noting that at $\sigma = 2$ the universal value already deviates from 0, indicating that the universality has changed. The gray bands mark the estimated critical point (vertical) and the extrapolated convergence value (horizontal). Similar behavior is observed for $Q_m$ in panels ($\rm a_2$), ($\rm b_2$), and ($\rm c_2$).}
    \label{fig:Overview}
\end{figure*}

As a precaution against correction-to-scaling terms that we missed including in the fitting ansatz, we impose a lower cutoff $L \ge L_{\rm min}$ on the data points admitted in the fit and systematically study the effect on the residuals $\chi^2$ value by increasing $L_{\rm min}$. In general, the preferred fit for any given ansatz corresponds to the smallest $L_{\rm min}$ for which the goodness of the fit is reasonable and for which subsequent increases in $L_{\rm min}$ do not cause the $\chi^2$ value to drop by vastly more than one unit per degree of freedom. In practice, by “reasonable” we mean that $\chi^2/\rm{DF} \approx 1$, where DF is the number of degrees of freedom. The systematic error is estimated by comparing estimates from various sensible fitting ansatz.

\begin{table*}[!ht]
    \centering
    \caption{Fitting results for critical polynomial $R_p$, using the ansatz given by Eq.~\eqref{eq:beta_c_fiting}.}
    \label{tab:Rp}
    \begin{tabular}{l|lllllllll}
    \hline\hline
        ~~~$\sigma$ & $L_{\rm min}$ & ~~~~~~~~$\beta_c$ & ~~~$y_t$ & ~~~$R_{p,c}$ & ~~~$a_1$ & ~~~$a_2$ & ~~~~$b_1$ & ~$y_1$ & $\chi^2$/DF \\ \hline
        1.75 & 256 & 0.329 136 1(1) & 1.011(8) & -0.326(3) & -3.8(2) & ~0.8(4) & -1.02(1)  & 0.3 & 28.6/26  \\ 
        ~ & 512 & 0.329 136 2(2) & 1.02(1) & -0.323(8) & -3.6(3) & ~0.6(4) & -1.05(5)  & 0.3 & 22.1/19  \\ 
        ~ & 256 & 0.329 135 6(2) & 1.008(9) & -0.370(2) & -3.8(2) & ~1.1(4) & -1.38(2)  & 0.4 & 31.2/26  \\ 
        ~ & 512 & 0.329 135 9(2) & 1.02(1) & -0.361(6) & -3.5(3) & ~0.7(4) & -1.50(8)  & 0.4 & 22.9/19  \\ \hline
        1.875 & 256 & 0.336 985 0(2)&	0.992(9)&	-0.185(4)&	-4.0(2)&	-1.3(4)	&-2.44(8)  & 0.5 & 17.7/19  \\ 
        ~ & 512 & 0.336 984 9(3)	& 0.99(1) &	-0.188(7) &	-4.1(3) &-1.3(4) &-2.4(1)  & 0.5 & 13.6/14  \\ 
        ~ & 256 & 0.336 984 7(2) &	0.991(9)&	-0.204(3)&	-4.0(2)&	-1.3(3)&	-3.7(1)  & 0.6 & 17.7/19  \\ 
        ~ & 512 & 0.336 984 7(3)&	0.99(1)&	-0.203(6)&	-4.1(3)&	-1.3(4)&	-3.8(2) & 0.6 & 13.5/14  \\ \hline
        2 & 256 & 0.344 439 5(2) & 1.002(7) & -0.058(2) & -3.5(1) & -1.3(5) & -1.36(2)  & 0.4 & 20.3/16  \\ 
        ~ & 512 & 0.344 439 5(3) & 1.011(9) & -0.057(4) & -3.3(2) & -0.9(5) & -1.37(5)  & 0.4 & 14.6/11  \\ 
        ~ & 256 & 0.344 438 9(2) & 1.005(7) & -0.084(2) & -3.4(1) & -1.3(6) & -1.96(3)  & 0.5 & 24.4/16  \\ 
        ~ & 512 & 0.344 439 2(3) & 1.011(9) & -0.079(4) & -3.2(2) & -0.9(5) & -2.08(8)  & 0.5 & 14.5/11  \\ \hline
        2.2 & 256 & 0.355 388 3(3) &0.99(1) & -0.006(2) & -3.5(3) &-2(1) &	-1.28(4) & 0.5	& 15.6/13   \\ 
        ~ & 512 & 0.355 388 5(4)  & 1.00(1)	&-0.004(4)	&-3.3(3)	&-1.6(9) & -1.3(1) &  0.5 & 10.4/11  \\ 
        ~ & 256 & 0.355 388 8(3) & 0.99(1) &	~0.011(3) &	-3.5(3) &	-2(1) &	-0.89(3) &  0.6 &	15.2/13  \\ 
        ~ & 512 & 0.355 388 8(4) & 1.00(1) &	~0.011(6) &	-3.3(3) &	-1.6(9) &-0.88(6) & 0.6 & 10.5/11  \\ \hline
        2.5 & 256 & 0.369 445 6(4) & 0.98(1) & ~0.004(2) & -3.4(2) & -4.2(9) & -0.72(7)  & 0.6 & 14.0/17  \\ 
        ~ & 512 & 0.369 446 1(5) & 0.98(1) & ~0.010(5) & -3.3(2) & -4.2(9) & -1.0(1)  & 0.6 & ~7.1/12  \\ 
        ~ & 256 & 0.369 445 4(4) & 0.98(1) & ~0.001(2) & -3.4(2) & -4.2(9) & -1.1(1)  & 0.7 & 14.4/17  \\ 
        ~ & 512 & 0.369 446 0(5) & 0.99(1) & ~0.007(4) & -3.3(2) & -4.2(9) & -1.5(3)  & 0.7 & ~7.1/12  \\
        \hline\hline
    \end{tabular}
\end{table*}

Here, we fit $R_p$ for $\sigma \le 2$ using Eq.~\eqref{eq:beta_c_fiting}. Firstly, we attempt to fit by setting $l = 2, 3$ and leaving all other parameters free, but this yields unstable results. Then, we fix $b_2 = c_1 = 0$, keeping only one correction term. We fix the correction exponent $y_1$ in the range $0.2$ to $1$, which produces relatively stable results. However, the $\chi^2/\mathrm{DF}$ value remains large, and the fitting results still vary significantly as $L_{\min}$ increases, indicating that the fit has not yet converged with increasing system size. 

When we set $L_{\min} \ge 256$, the $\chi^2/\mathrm{DF}$ value gradually approaches $1$, and further increasing $L_{\min}$ leads to only small variations in the coefficient $q$, suggesting that we have achieved a good fit. Since the correction form in the long-range case is not well established, we tested several different correction schemes, and the results are summarized in Table~\ref{tab:Rp}.

From the Table, we can see that all the critical points are consistent with the extrapolated results in the main text within the error bars, supporting our estimation of the critical points. Through the fitting, we also obtain the critical exponent $y_t$, which is approximately $1$ in all cases. Therefore, it is difficult to distinguish the universality class from $y_t$, consistent with the conclusions in the main text. For the universal value, we find that it agrees with the extrapolated result, though it shows noticeable variations depending on the correction scheme. Considering the systematic errors, we obtain $R_{p,c} = -0.34(3), -0.20(2), -0.07(2)$ for $\sigma = 1.75, 1.875,$ and $2$, respectively. Overall, although the fitted results show larger uncertainties compared with the extrapolated ones in the main text, the central values remain consistent with the extrapolations.

\begin{table*}[!ht]
    \centering
    \caption{Fitting results for Binder ratio $Q_m$, using the ansatz given by Eq.~\eqref{eq:beta_c_fiting}.}
    \label{tab:Qm}
    \begin{tabular}{l|lllllllll}
    \hline\hline
        ~~~$\sigma$ & $L_{\rm min}$ & ~~~~~~~~$\beta_c$ & ~~~$y_t$ & ~~~$Q_{m,c}$ & ~~~$a_1$ & ~~~$a_2$ & ~~~~$b_1$ & ~$y_1$ & $\chi^2$/DF \\ \hline
        1.75 & 256 & 0.329 136 6(2) & 1.00(1) & 0.804(1) &-0.74(6) & -0.3(1) &-0.303(4) & 0.25 & 35.8/26  \\ 
        ~ & 512 & 0.329 136 6(3) &1.01(1)	& 0.805(2) &-0.67(8) & -0.3(1)	& -0.31(1)  & 0.25 & 26.7/19 \\ 
        ~ & 256 & 0.329 136 1(2) &	1.00(1)	& 0.792 5(9) & -0.75(6)&	-0.3(1)&-0.340(4) & 0.3 & 35.8/26  \\ 
        ~ & 512 & 0.329 136 3(3)	&1.01(1) & 0.795(2) &-0.67(8) &	-0.2(1) &-0.35(1)  & 0.3 & 26.7/19 \\  \hline
        1.875 & 256 & 0.336 984 3(2) & 0.98(1) & 0.812 8(7) &	-0.74(5) & -0.7(1) &-1.11(2)  & 0.5 & 13.8/19  \\ 
        ~ & 512 & 0.336 984 4(3) & 0.98(1) & 0.814(1)	& -0.74(5) & -0.7(1) & -1.15(5)  & 0.5 & ~8.4/14 \\ 
        ~ & 256 & 0.336 984 8(3) & 0.98(1) &	0.818 4(8) &	-0.74(5) &	-0.7(1)	& -0.72(1)  & 0.6 & 13.8/19  \\ 
        ~ & 512 & 0.336 984 8(3) & 0.98(1) &	0.818(1)  &	-0.75(6) &	-0.7(1)	& -0.71(3) & 0.6 & ~8.9/14  \\ \hline
        2 & 256 & 0.344 440 9(3) & 0.98(1) &	0.859 5(7) &	-0.65(4) & -0.7(2) & -0.286(4)  & 0.3 & 23.3/16  \\ 
        ~ & 512 & 0.344 440 4(4) & 1.00(1) &	0.857(1)  &	-0.60(4) & -0.5(2) & -0.271(8)  & 0.3 & 15.3/11  \\ 
        ~ & 256 & 0.344 439 8(3) & 0.990(9) &	0.847 2(5) &	-0.62(3) &	-0.7(1) &	-0.388(4)  & 0.4 & 19.1/16  \\ 
        ~ & 512 & 0.344 439 7(4) & 1.00(1)  &	0.847(1)  &	-0.59(4) &	-0.5(1)	& -0.39(1)  & 0.4 & 15.3/11  \\ \hline
        2.2 & 256 & 0.355 388 6(4) & 0.99(1) & 0.855 9(5) & -0.56(6) &	-0.6(2) & -0.363(9)  & 0.5 & 13.0/13  \\ 
        ~ & 512 & 0.355 389 2(4) & 0.99(1) &	0.857 4(8) &	-0.55(5) & -0.6(2) & -0.40(1)  & 0.5 & ~7.6/11  \\ 
        ~ & 256 & 0.355 387 8(4) & 0.99(1) & 0.852 7(5) & -0.55(7)	& -0.6(3) &	-0.54(1)  & 0.6 & 16.7/13  \\ 
        ~ & 512 & 0.355 388 7(4) & 0.99(1) & 0.854 5(7) & -0.55(5) & -0.6(2) &	-0.63(2)  & 0.6 & ~7.4/11 \\ \hline
        2.5 & 256 & 0.369 445 2(6) & 0.97(1) & 0.857 2(6) & -0.58(6) &	-0.9(2) &-0.22(1)  & 0.6 & 16.7/17  \\ 
        ~ & 512 & 0.369 445 4(8) & 0.98(1) &	0.857(1) &	-0.54(7) & -0.8(2) & -0.23(4)  & 0.6 & ~9.8/12  \\ 
        ~ & 256 & 0.369 444 8(6) & 0.97(1) &	0.856 1(5) &	-0.58(6) &	-0.9(2) & -0.34(2)  & 0.7 & 17.0/17  \\ 
        ~ & 512 & 0.369 445 2(8) & 0.98(1) &	0.857(1) &	-0.54(7) &	-0.8(2)	& -0.38(7)  & 0.7 & ~9.8/12  \\ 
        \hline\hline
    \end{tabular}
\end{table*}

The fitting process for $Q_m$ is similar, and the final results are summarized in Table~\ref{tab:Qm}. We can see that the corrections to $Q_m$ are also significant, which requires setting $L_{\min} \ge 256$. However, compared with $R_p$, the deviation of the universal value of $Q_m$ from the SR case is much smaller. Specifically, at $\sigma = 1.75$, the gap is about $0.34$ for $R_p$ but less than $0.1$ for $Q_m$; at $\sigma = 2$, the contrast is even more pronounced, with the gap being about $0.06$ for $R_p$ but less than $0.01$ for $Q_m$. Therefore, the difference between universality classes tends to be obscured by the error bars. In addition, $Q_m$ yields critical points consistent with those obtained from $R_p$.

\begin{figure}
    \centering
    \includegraphics[width=\linewidth]{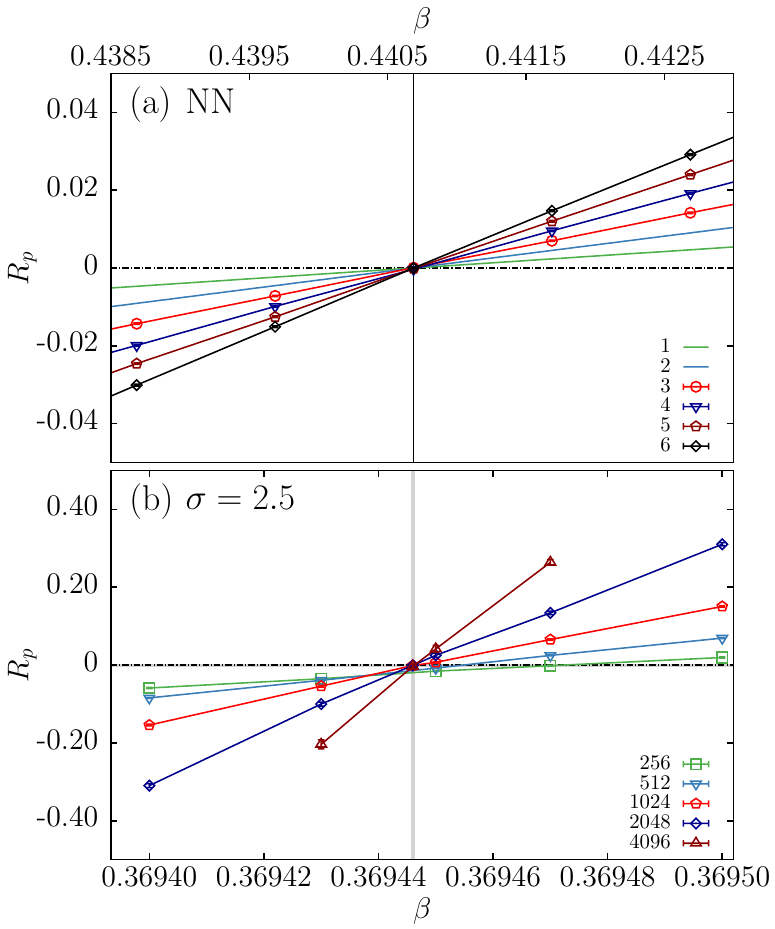}
    \caption{Demonstration of the critical polynomial $R_p$ for the 2D SR Ising universality. Both the NN and $\sigma=2.5$ cases converge to zero at criticality. The solid vertical line marks the critical point for NN, and the two gray bands indicate the estimated critical point (vertical) and convergence value (horizontal) for $\sigma=2.5$.}
    \label{fig:Rp_SR}
\end{figure}

Besides the $\sigma \leq 2$ cases, we also perform fits for $\sigma = 2.2$ and $\sigma = 2.5$, and the corresponding results are summarized in Table~\ref{tab:Rp} and Table~\ref{tab:Qm}. We can see that the critical points obtained from these fits are consistent, within the error bars, with the extrapolated results presented in the main text. Moreover, within twice the error range, both $R_p$ and $Q_m$ include their respective SR universal values, further confirming the robustness of our fitting procedure.

Although the properties of the SR case are well known, for the sake of numerical completeness, it is worth showing the $R_p$ for $\sigma > 2$ in Fig.~\ref{fig:Rp_SR}. For the NN case in Fig.~\ref{fig:Rp_SR}(a), following the definition of $R_p$, we can obtain the exact expressions for $L=1,2$ as follows:
\begin{align}
R_p(\beta, L=1) &= 1 - e^{-4\beta} - 2 e^{-2\beta}, \nonumber \\
R_p(\beta, L=2) &= 
\frac{(1 + e^{4\beta})^2 (1 - 6 e^{4\beta} + e^{8\beta})}
     {1 + 6 e^{8\beta} + e^{16\beta}}. \nonumber
\end{align}
These two curves are plotted in Fig.~\ref{fig:Rp_SR}(a), where the green line corresponds to $L=1$ and the blue line to $L=2$. For $L=3,4,5,6$, we provide the MC data. We can see that, for the NN case, there is no finite-size correction; all crossing points fall exactly on $(\beta_c, 0)$. When a long-range interaction is introduced but remains weak, for example, $\sigma = 2.5$ as shown in Fig.~\ref{fig:Rp_SR}(b), finite-size corrections appear but converge rapidly to $(\beta_c, 0)$ for $L > 256$. The gray bands in the figure represent our fitted results.

\renewcommand{\thefigure}{B\arabic{figure}}
\renewcommand{\thetable}{B\arabic{table}}
\setcounter{figure}{0}
\setcounter{table}{0}

\renewcommand{\theHequation}{B\arabic{equation}}
\renewcommand{\theHfigure}{B\arabic{figure}}
\renewcommand{\theHtable}{B\arabic{table}}

\section{least-squares fits of the critical exponents}

To extract the critical exponents $\eta$ and $y_t$, we fit the observables to the finite-size scaling form  
\begin{align}
    \mathcal{O} \;=\; L^{y_{\mathcal{O}}}\big(a_0 + b_1 L^{-y_1} + b_2 L^{-y_2}\big) + c,
    \label{eq:scaling_fitting}
\end{align}
with $\mathcal{O} = \chi,\, C_1,\, g_{ER}^{(x)}$ as appropriate for each measurement. Here, $y_{\mathcal{O}}$ denotes the leading scaling exponent of observable $\mathcal{O}$, and the terms $b_i L^{-y_i}$ account for the leading finite-size corrections. The parameter $c$ originates from the analytic part of the free energy.

To extract the anomalous dimension $\eta$, we focus on the susceptibility $\chi \sim L^{2-\eta}$ and the largest cluster $C_1 \sim L^{2-\eta/2}$ in the FK representation. Initially, we keep the fitting parameter $c$ free, but regardless of how other parameters are adjusted, the uncertainty of $c$ remains large. Therefore, we later fix $c = 0$ in the fits. In this case, the $\chi^2/{\rm DF}$ is close to 1, the uncertainties of each parameter are reasonable, and the fitting results remain stable when increasing $L_{\min}$. The results are summarized in Table~\ref{tab:chi} for $\chi$ and in Table~\ref{tab:C1} for $C_1$.

\begin{table}[!ht]
    \centering
    \caption{Fitting results for $\chi$, using the ansatz given by Eq.~\eqref{eq:scaling_fitting}.}
    \label{tab:chi}
    \begin{tabular}{l|llllll}
    \hline\hline
        $\sigma$ & $L_{\rm min}$ & ~~~~~$\eta$ & ~~~$a_0$ & ~~~$b_1$ & ~~~$y_1$ & $\chi^2$/DF  \\ \hline
        1.75 & 16	& 0.335(1) &  0.459(5) & 0.811(6) & 0.577(8) & 6.7/6 \\
        ~ & 32  & 0.336(2)	&  0.461(9) & 0.82(2)  & 0.58(2)  & 6.6/5 \\ \hline
        1.875 & 16 &0.293(1) &	0.497(5) & 0.734(9) & 0.62(1) & 5.3/6  \\ 
        ~ & 32 & 0.293(2) & 0.498(9) &	0.74(3)	& 0.62(2) & 5.3/5 \\ \hline
        2 & 64 & 0.274(1) & 0.571(7) &	0.8(1) & 0.76(6) &	4.6/4  \\ 
        ~ & 128 & 0.273(4) & 0.57(2) &	0.7(4) & 0.7(2) & 4.4/3 \\ \hline
        2.5 & 16 & 0.2501(1) & 0.790(1) & 0.308(9) & 0.84(1) & 1.2/5  \\ 
        ~ & 32 & 0.2497(3) & 0.788(1) & 0.27(2) & 0.79(4) & 0.8/4  \\ \hline\hline
    \end{tabular}
\end{table}

\begin{table}[!ht]
    \centering
    \caption{Fitting results for $C_1$, using the ansatz given by Eq.~\eqref{eq:scaling_fitting}.}
    \label{tab:C1}
    \begin{tabular}{l|llllll}
    \hline\hline
        $\sigma$ & $L_{\rm min}$ & ~~~~~$\eta/2$ & ~~~$a_0$ & ~~~$b_1$ & ~~~$y_1$ & $\chi^2$/DF  \\ \hline
        1.75 & 32 & 0.166(1) &	0.627(5) &	0.45(1) & 0.61(2) & 5.3/5  \\ 
        ~ & 64 & 0.165(2) & 0.62(1) &	0.42(6) & 0.58(8) & 5.1/4  \\ \hline
        1.875 & 16 & 0.1453(7)	& 0.664(3) & 0.40(1) &	0.68(2) &	6.1/6  \\ 
        ~ & 32 & 0.145(1) & 0.664(6) &	0.40(4)	& 0.69(5) & 6.1/5 \\ \hline
        2 & 64 & 0.1361(6)	& 0.719(3) & 0.8(4) & 1.1(1) &	5.8/4  \\ 
        ~ & 128 & 0.136(2)	& 0.72(1) &	0.4(7) & 0.8(6) &	5.6/3  \\ \hline
        2.5 & 8 & 0.12509(4) & 0.8568(1) & 0.32(2) & 1.68(3) & 2.5/6  \\ 
        ~ & 16 & 0.12503(6) & 0.8565(3) & 0.21(6) & 1.5(1) & 1.9/5  \\ \hline\hline
    \end{tabular}
\end{table}

We find that the values of $\eta$ extracted from $\chi$ and $C_1$ are consistent within the error bars. It is evident that for $\sigma = 1.75$ and $1.875$, $\eta$ deviates significantly from the short-range value $1/4$. By taking the systematic errors into account, we estimate $\eta = 0.335(4)$ for $\sigma = 1.75$ and $\eta = 0.293(3)$ for $\sigma = 1.875$. At $\sigma = 2$, a slight deviation remains, which is consistent and self-consistent with our previous analysis of the universal values. Considering the systematic errors, we estimate $\eta = 0.273(3)$. For $\sigma = 2.5$, the results return to the short-range value $1/4$, and we obtain $\eta = 0.250(1)$.

\begin{figure*}
    \centering
    \includegraphics[width=\linewidth]{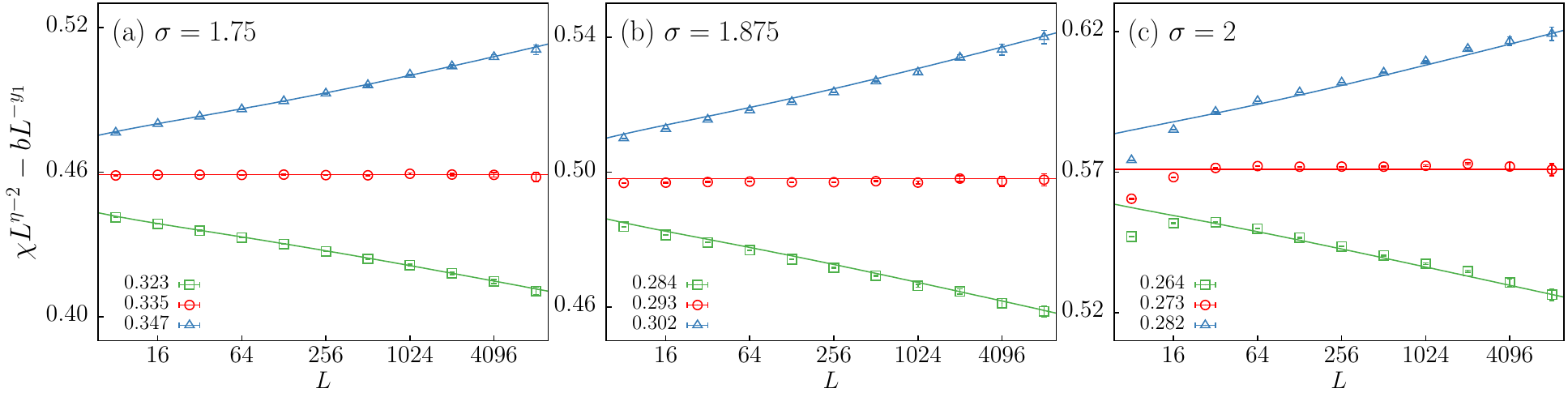}
    \caption{Demonstration of the reliability of the fit of $\chi$.  Semilogarithmic plots of $\chi L^{\eta - 2} - bL^{-y_1}$ versus $L$ are shown for $\sigma = 1.75$, $1.875$, and $2$, with $y_1 = 0.577$, $0.62$, and $0.76$, respectively. The red dots correspond to the fitted central value $\eta_c$, while the green and blue dots represent $\eta$ values deviating by positive and negative three standard deviations, respectively. 
    The red points approach a horizontal line as $L \to \infty$, whereas the others exhibit clear upward or downward trends, demonstrating the high precision of our estimate for $\eta$.}
    \label{fig:eta_precise}
\end{figure*}

To further demonstrate the stability of the fits, we show the results in Fig.~\ref{fig:eta_precise}. Specifically, when using the estimated value of $\eta$, according to the fitting relation, $\chi L^{\eta - 2} - bL^{-y_1}$ should converge to a nonzero constant as $L$ increases; otherwise, it will either diverge or vanish. As shown in Fig.~\ref{fig:eta_precise}, the curves corresponding to the estimated exponents converge to a plateau, indicating the reliability of the fitting results.

As for the thermal scaling exponent $y_t = 1/\nu$, where $\nu$ is the correlation-length exponent, we fit the covariance $g^{(x)}_{ER}$. The fitting process is similar, and we again find that setting $c = 0$ yields better results. Moreover, since $g^{(x)}_{ER}$ is a higher-order observable, the corrections are significant, requiring a larger $L_{\min}$ than in the case of fitting $\eta$. Our results are summarized in Table~\ref{tab:gx}.

\begin{figure*}
    \centering
    \includegraphics[width=\linewidth]{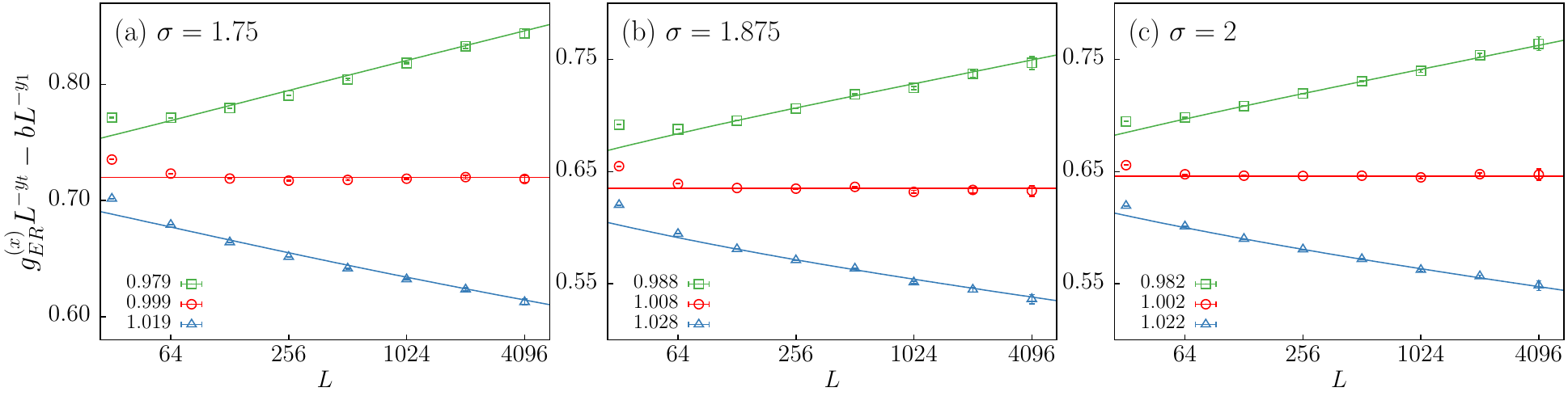}
    \caption{Demonstration of the reliability of the fit of $g^{(x)}_{ER}$. Semilogarithmic plots of $g^{(x)}_{ER} L^{-y_t} - bL^{-y_1}$ versus $L$ are shown for $\sigma = 1.75$, $1.875$, and $2$, with $y_1 = 0.45$, $0.8$, and $0.8$, respectively. The red dots correspond to the fitted central value $y_t$, while the green and blue dots represent $y_t$ values deviating by positive and negative three standard deviations, respectively. 
    The red points approach a horizontal line as $L \to \infty$, whereas the others exhibit clear upward or downward trends, demonstrating the high precision of our estimate for $y_t$.}
    \label{fig:yt_precise}
\end{figure*}

We find that $y_t$ is close to $1$ for all fitted values of $\sigma$, indicating that this exponent is not suitable for distinguishing between different universality classes. Here, it is shown only for the completeness of our analysis. 
In addition, similar to the case of $\eta$, we plot Fig.~\ref{fig:yt_precise} to illustrate the reliability of the fitting results.

\begin{table}[!ht]
    \centering
    \caption{Fitting results for $g^{(x)}_{ER}$, using the ansatz given by Eq.~\eqref{eq:scaling_fitting}.}
    \label{tab:gx}
    \begin{tabular}{l|llllll}
    \hline\hline
        $\sigma$ & $L_{\rm min}$ & ~~~~~$y_t$ & ~~~$a_0$ & ~~~$b_1$ & ~$y_1$ & $\chi^2$/DF  \\ \hline
        1.75 & 64 &	1.003(1) & 0.707(8)	& -0.85(2) & 0.4 &	4.3/5  \\ 
        ~ & 128	& 1.003(3) &	0.71(1)	& -0.84(6) & 0.4 &	4.3/4  \\
        ~ & 256	& 1.007(6) & 0.66(3) &	-1.7(3)	& 0.6	& 3.3/3  \\
        ~ & 512	& 0.990(9) & 0.76(6) &	-3.2(9)	& 0.6	& 1.2/2  \\ \hline
        1.875 & 256 &	0.991(6) &	0.73(3) &	-1.8(3)	& 0.6 &	3.5/3  \\ 
        ~ & 512	& 1.008(9) &	0.63(5) &	-0.3(6) & 0.6 &	1.1/2 \\
        ~ & 256	& 0.997(4) &	0.69(2) &	-3.6(6)	& 0.8 &    2.9/3 \\
        ~ & 512	& 1.010(8) &	0.62(4)	&   -1(1)	& 0.8 &	1.2/2\\ \hline
        2 & 128	& 0.996(2) & 0.68(1)  &	-1.00(9) & 0.6	& 5.1/4 \\
        ~& 256	& 0.998(5) & 0.67(3)  &	-0.9(2)	 & 0.6	& 4.9/3 \\
        ~ & 128	& 1.002(2) & 0.645(9) &	-1.8(1)	 & 0.8	& 4.8/4 \\
        ~ & 256	& 1.002(4) & 0.65(2)  &	-1.8(5)	 & 0.8	& 4.8/3\\ \hline
        2.5 & 64 &	1.005(3) &	0.62(1) &	-0.63(6) &	0.59(5) &	0.4/3  \\ 
        ~ & 128	& 1.009(4) & 0.60(1) & -0.9(4) &	0.7(1)	& 0.3/2  \\ \hline\hline        
    \end{tabular}
\end{table}

\bibliography{ref}

\end{document}